\documentclass[journal,transmag]{IEEEtran}
\usepackage{times}

\usepackage{soul}
\usepackage{url}
\usepackage[utf8]{inputenc}
\usepackage{graphicx}
\usepackage{epstopdf}
\usepackage{amsmath}
\usepackage{amssymb}
\usepackage{amsthm}
\usepackage{booktabs}
\usepackage{algorithm}
\usepackage{algorithmicx}
\usepackage{algpseudocode}
\usepackage{amsfonts}
\usepackage{bm}
\usepackage{subfigure}
\usepackage{multirow}
\usepackage{longtable}
\usepackage{color}
\usepackage{textcomp}
\usepackage[figuresright]{rotating}
\usepackage{blindtext}
\usepackage[colorlinks=true,linkcolor=black,citecolor=black,urlcolor=black,]{hyperref}
\usepackage[doipre={doi:~}]{uri}
\usepackage{setspace}
\usepackage{indentfirst}
\usepackage{bigstrut}
\usepackage{tabularx}
\usepackage{tabu}
\usepackage{lineno}
\let\oldequation\equation
\let\oldendequation\endequation
\usepackage{fancyhdr}
\renewenvironment{equation}
  {\linenomathNonumbers\oldequation}
  {\oldendequation\endlinenomath}

\begin{document}
\fancyhead[C]{This article is used as the preliminary studies for Lee Kuan Yew Postdoctoral Fellowship (LKYPDF) in Singapore}
\begin{sloppypar}
\title{An open unified deep graph learning framework for discovering drug leads}

\author{\IEEEauthorblockN{
Yueming Yin,
Haifeng Hu,
Zhen Yang,
Jitao Yang,
Chun Ye,
Jiansheng Wu, and
Wilson Wen Bin Goh}
\thanks{
Yueming Yin, Haifeng Hu, Zhen Yang, Jitao Yang and Chun Ye are with the School of Telecommunications and Information Engineering, Nanjing University of Posts and Telecommunications, Nanjing 210003, China.

Yueming Yin is with the School of Computer Science and Engineering, Nanyang Technological University, 637551, Singapore

Zhen Yang is with the National Engineering Research Center of Communications and Networking, Nanjing University of Posts and Telecommunications, Nanjing 210003, China.

Jiansheng Wu is with the School of Geographic and Biologic Information, and with Smart Health Big Data Analysis and Location Services Engineering Research Center of Jiangsu Province, Nanjing University of Posts and Telecommunications, Nanjing 210023, China.

Wilson Wen Bin Goh is with Lee Kong Chian School of Medicine, School of Biological Sciences, Nanyang Technological University, 637551, Singapore, and Center for Biomedical Informatics, 636921, Singapore.

Corresponding author: Yueming Yin (yinym96@qq.com), Haifeng Hu (huhf@njupt.edu.cn), Jiansheng Wu (jansen@njupt.edu.cn) and Wilson Wen Bin Goh (wilsongoh@ntu.edu.sg).}}

\maketitle

\begin{abstract}
Computational discovery of ideal lead compounds is a critical process for modern drug discovery. It comprises multiple stages: hit screening, molecular property prediction, and molecule optimization. Current efforts are disparate, involving the establishment of models for each stage, followed by multi-stage multi-model integration. However, this is non-ideal, as clumsy integration of incompatible models increases research overheads, and may even reduce success rates in drug discovery. Facilitating compatibilities requires establishing inherent model consistencies across lead discovery stages. Towards that effect, we propose an open deep graph learning (DGL) based pipeline: generative adversarial feature subspace enhancement (GAFSE), which first unifies the modeling of these stages into one learning framework. GAFSE also offers standardized modular design and streamlined interfaces for future expansions and community support. GAFSE combines adversarial/generative learning, graph attention network, graph reconstruction network, and optimizes the classification/regression loss, adversarial/generative loss, and reconstruction loss simultaneously. Convergence analysis theoretically guarantees model generalization performance. Exhaustive benchmarking demonstrates that the GAFSE pipeline achieves excellent performance across almost all lead discovery stages, while also providing valuable model interpretability. Hence, we believe this tool will enhance the efficiency and productivity of drug discovery researchers.
\end{abstract}


\section{Introduction}
\IEEEPARstart{C}{omputational} discovery of ideal lead compounds is a critical process in modern drug discovery \cite{ou2012computational,shaker2021silico}. This involves multiple stages, including hit screening, molecular property prediction, and molecule optimization. Currently, given the availability of big data, a common approach for hit screening is to represent compound molecules as a graph structure, train a deep graph learning model and screen hits according to their predicted bioactivities \cite{bahi2018deep}. In addition to bioactivity, accurate drug discovery entails multiple molecular property prediction problems spanning multiple length scales, including physicochemical, geometric, energetic, electronic, and thermodynamic properties \cite{ramakrishnan2014quantum}. Given such high information modalities and complexities, training deep graph learning to predict molecular properties is a convenient and consequently, popular approach \cite{wieder2020compact}. As a critical step in drug development, molecule optimization improves the desired properties of drug candidates through chemical modification. It includes the optimization of hits to leads, and the optimization of leads towards viable drug development. Currently, a popular practice is to predict potential alternative sites by deep graph learning on given molecular graphs, and perform removal and/or addition of atoms or fragments at that site. It typically includes optimizing the binding activity of hits and improving the ADMET properties of leads.

\begin{figure}[t]
\centering
\includegraphics[width=1\linewidth,height=1.5in]{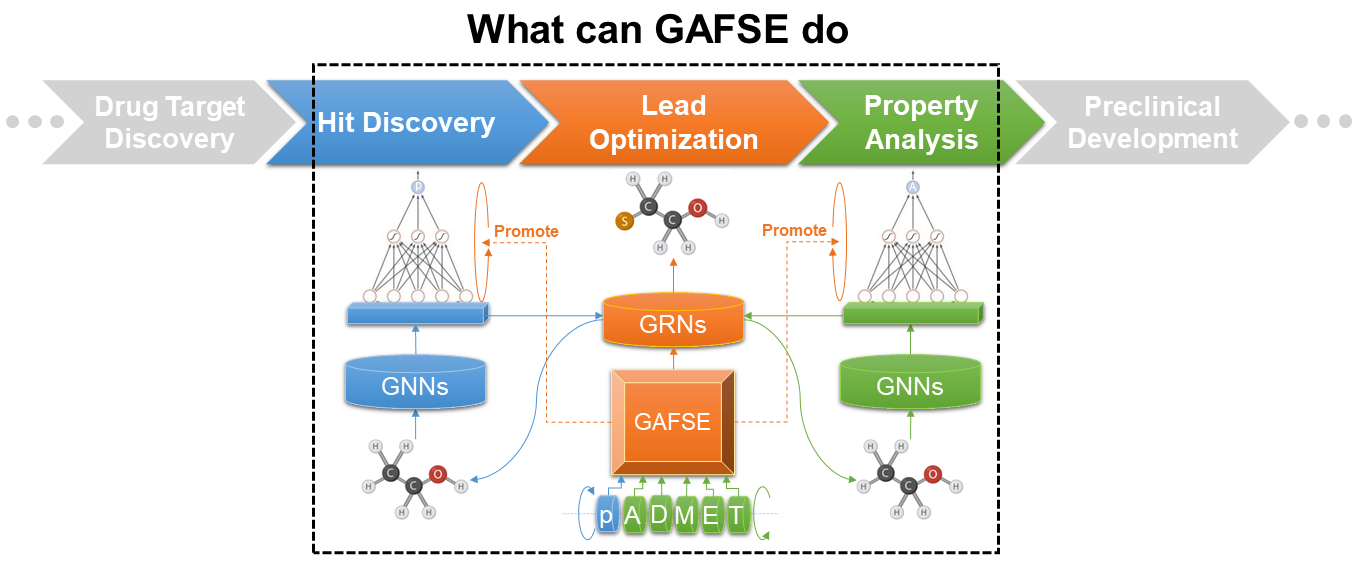}
\caption{Illustration of the proposed GAFSE framework.}
\label{framework}
\end{figure}

Current computational drug discovery scheme is to build discrete models for each stage followed by model integration \cite{shaker2021silico}. However, clumsy comparisons across incompatible models may hamper efficiency and reduce success rates. These discrete models were developed based on different frameworks, different data, or even different computational languages. Thus, how to select and use these disparate models (for each process) becomes a challenging problem in itself. Moreover, this also leads towards problems of consistency and synergy between the multitude of machine learning models used across the myriad processes of lead discovery \cite{xiao2022synergy}. Our previous research found that the modeling of many processes in lead discovery was consistent in nature and could be better exploited \cite{wu2018wdl,wu2019precise,wu2020homologous,yin2021realvs,yin2022afse}. Firstly, the objects of these processes are small compound molecules, which can be naturally represented as graph structures (deep graph learning methods can naturally be developed to model these processes). Secondly, the problems of these process studies can be abstracted into the prediction of property values of compound molecules, that is, the quantitative structure-property relationship (QSPR) problem \cite{le2012quantitative}. For hit screening, we predict the relationship between compound structure and activity; for molecular property prediction, we study the relationship between compound structure and various molecular properties; for molecule optimization, we study the relationship between compound structure optimization and molecular activity and/or ADMET properties. These related objectives can be unified.

Therefore, we propose a unified learning framework to achieve better model coherence. Interestingly, despite the need, there are no other reported works in this area. Our unified learning framework is meant to: (1) Resolve difficulties in user use and secondary development; (2) Facilitate model evaluation and selection; (3) Improve model coherence across multiple processes; (4) Improve the success rate of lead discovery. Our unified deep graph learning framework is open. This brings the following benefits: (1) facilitate community participation in solution building across the myriad of lead compound discovery processes; (2) facilitate code reusability, model reproducibility and transparency via the adoption of standardized modules, streamlined interfaces, and detailed documentation.

However, unifying hit screening, molecular property prediction and molecule optimization in one framework will bring many challenges, because each of these processes has its unique characteristics. For example, some tasks are better resolved via classification methods while others are by regression learning. Within themselves, there are problems of small samples and unbalanced samples in classification learning. And in regression problems, there will be problems of activity cliffs, inconsistent distribution of training and testing samples, and small samples. Molecular optimization problems take into account the bioactivity and specificity of molecules, and thus need to optimize molecules for multiple goals. To build an open unified deep graph learning framework for discovering drug leads while also considering the specificity of these steps, this paper considers the following aspects: (1) Objective function. Various loss functions are introduced in the framework, including classification/regression/adversarial/generation, etc. (2) Small samples. Semi-supervised learning and multi-task learning are introduced in the framework. (3) Activity cliffs. Adversarial learning and generative learning are introduced in the framework. (4) Molecular optimization. In the framework, graph attention mechanism, graph reconstruction network, and matching molecular pairs on activity cliffs (MMP-Cliffs) \cite{cruz2014activity} are introduced. Of course, the discovery of drug leads is complicated, and to solve it better, more learning methods need to be introduced in the future.

In this paper, we constructed an open unified deep graph learning framework GAFSE for discovering drug leads. For the screening of hit compounds, we develop the algorithm GAFSE-HS on the GAFSE framework, and its results on the GPCR benchmark dataset show that it exceeds the state-of-the-art bioactivity regression algorithm AFSE \cite{yin2022afse}. For molecular property prediction, we develop the algorithm GAFSE-MP on the GAFSE framework, and its results on the ADMET benchmark dataset show that it exceeds the state-of-the-arts molecular property prediction algorithm ADMETlab 2.0 \cite{xiong2021admetlab}. For molecule optimization, we developed the algorithm GAFSE-MO on the GAFSE framework. The results on the molecule optimization dataset generated by the above benchmark datasets show that GAFSE-MO obtains precise optimization results of molecular activities and properties; and the results on the COVID-19-related dataset (AID1706 Bioassay Data) show that, GAFSE-MO also outperforms the state-of-the-arts molecular generation algorithm GEOM-CVAE \cite{li2022geometry}. In addition, the convergence of the core adversarial algorithm in our learning framework GAFSE is theoretically proved, and an adaptive learning adjustment mechanism is designed accordingly.

\begin{figure*}[!h]
\centering
\includegraphics[width=0.9\linewidth,height=6in]{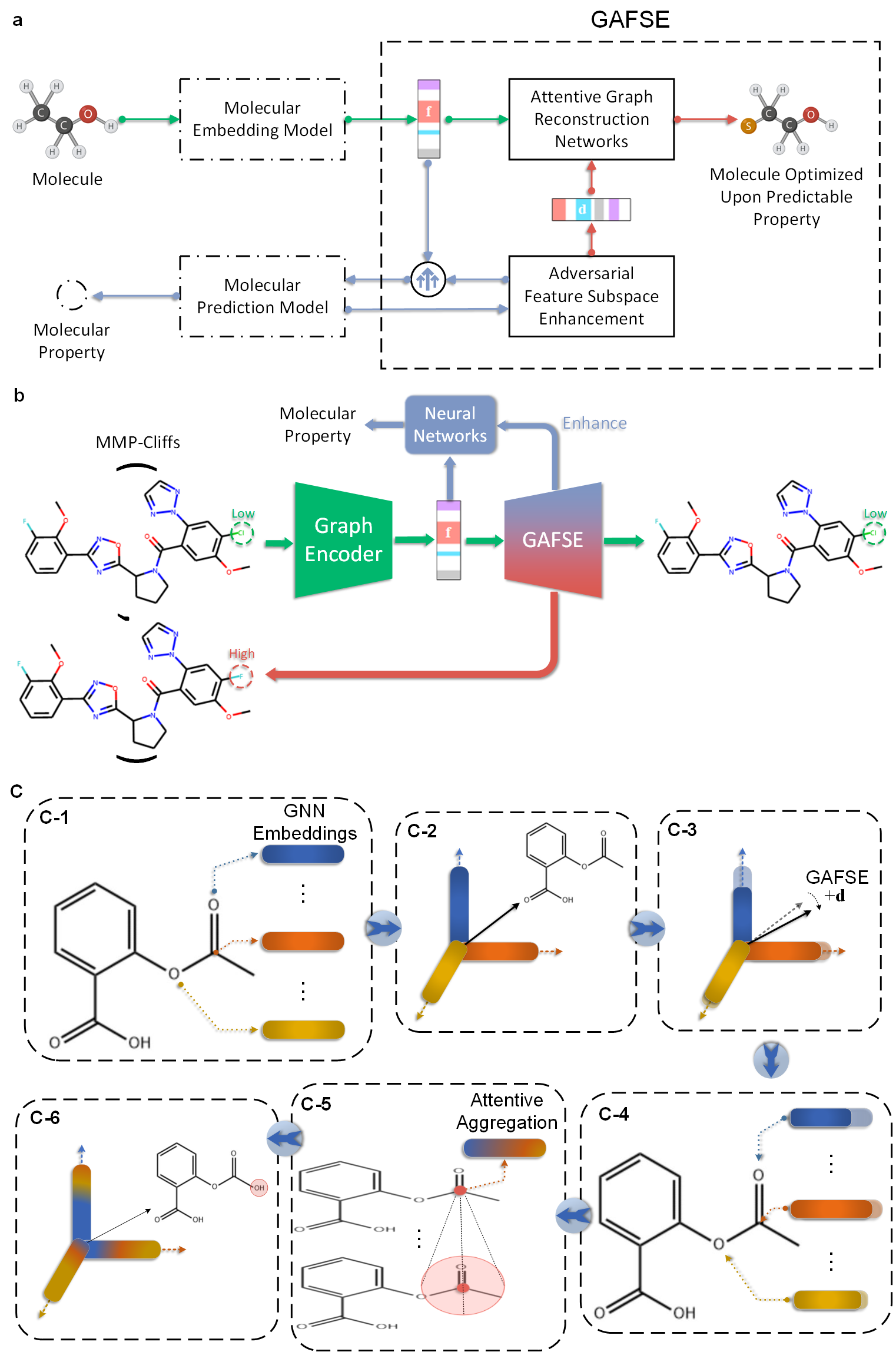}
\caption{a) Illustration of the proposed GAFSE framework. b) Illustration of generating MMP-Cliffs by the proposed GAFSE. “MMP-Cliffs” means matched molecular pairs, which are defined as a pair of molecules that differ only by a single chemical transformation. c) Schematic of GAFSE molecule optimization: (c-1) Graph embeddings of atoms. (c-2) Aggregation of atom embeddings and reconstruction of molecules. (c-3) The $d$ generated by GAFSE is mapped onto atom embeddings. (c-4) Updates to atom embeddings. (c-5) Graph attention aggregation among atom embeddings. (c-6) Modified position and element estimated from updated atom embeddings.}
\label{framework}
\end{figure*}

\section{Methods}
The framework of GAFSE is shown in Figure \ref{framework}a. To unify hit screening, molecular property prediction, and lead optimization openly, the GAFSE framework can be embedded between molecular features and downstream tasks in general molecular property prediction (MP) pipelines. By joining the GAFSE framework, general MP models will generate molecules with optimized properties while the model generalization is enhanced. Outside the GAFSE framework, the molecular embedding model (this paper takes Attentive FP \cite{xiong2019pushing} as an example, see supplementary Algorithm S1 for its implementation) produces molecular embeddings $\mathbf{f}$ to capture the key features of molecular graphs, and feeds them into the GAFSE framework to reconstruct the input molecules (Figure \ref{framework}a: green lines). Then, the MP model fits the function between molecular embeddings and their property values through multi-layer fully connected neural networks, and inputs its gradient on molecular embeddings into the GAFSE framework (Figure \ref{framework}a: the input blue line of GAFSE). Inside the GAFSE framework, the adversarial subspace enhancement algorithm (Section \ref{secGAFSE}-1: AFSE) generates adversarial perturbations $\mathbf{d}$ based on the gradient of molecular embeddings, and enhances the generalizability of the MP model together with a theoretical adaptive learning rate (Figure \ref{framework}a: the output blue line of GAFSE, detailed in Section \ref{secConvergence}). Meanwhile, the attentive graph reconstruction networks (Section \ref{secAGRN}: AGRNs) reconstruct the optimized molecule upon the predictable property according to the molecular embedding $\mathbf{f}$ and the adversarial perturbation $\mathbf{d}$ (Figure \ref{framework}a: red lines, detailed in Section \ref{secGAFSE}). See supplementary Table S1 for the notations used in this section.

\subsection{Attentive Graph Reconstruction Networks}
\label{secAGRN}
The graph attention mechanism \cite{vaswani2017attention} has been validated in \cite{xiong2019pushing} to have important utility and interpretability for targeted extraction of molecular embeddings. Its interpretability is reflected in the focus on key atoms \cite{yin2021realvs}. However, this is not the ultimate goal of drug discovery, and subsequent changes and optimizations on key atoms are almost the only way to develop molecular drugs. To predict possible optimization elements at key atomic positions, we propose a novel Attentive Graph Reconstruction Networks (AGRNs) to provide models for subsequent molecule optimization. 

The algorithm of AGRNs is shown in the supplementary Algorithm S2, which mainly includes the following key steps:\\
(1) Feature Mapping:
\begin{equation}
\begin{aligned}
\gamma_i=\text{softmax}(\langle\mathbf{f}, \mathbf{h}_i\rangle).
\label{eqMapping}
\end{aligned}
\end{equation}

To map the molecular embedding to its internal atoms, Eq. \ref{eqMapping} maps the information contained in the molecular embedding to each atom by conducting the inner product of the molecular embedding $\mathbf{f}$ and atomic embeddings $\mathbf{h}_i$. Where $\langle\cdot\rangle$ represents the vector inner product, and “softmax” represents the Softmax activation function. \\
(2) Feature Updating:
\begin{equation}
\begin{aligned}
\mathbf{g}_i=\gamma_i \mathbf{f}+\mathbf{h}_i.
\label{eqWeighting}
\end{aligned}
\end{equation}

To obtain the features of generated atoms, Eq. \ref{eqWeighting} updates the hidden layer features of each atom $\mathbf{h}_i$ by taking $\gamma_i$ as the step and along the direction of the molecular embedding $\mathbf{f}$.\\
(3) Feature Relating:
\begin{equation}
\begin{aligned}
\mathbf{r}_i=\text{elu}(\mathbf{W}\cdot(\text{dropout}([\mathbf{f},\mathbf{g}_i]))+\mathbf{c}).
\label{eqAttention}
\end{aligned}
\end{equation}

\begin{table*}[htbp]
\centering
\caption{Initial Atomic and Bond Features}
\label{tInitialF}
\resizebox{\textwidth}{28mm}{
\begin{tabular}{cccccc}
\hline
atom feature & type & size & description & $\phi_a$ & $L_a$ \\
\hline
atom symbol & one-hot & 16 & [B, C, N, O, F, Si, P, S, Cl, As, Se, Br, Te, I, At, metal] & Softmax & WCE \textsuperscript{\emph{a}} \\
degree & one-hot & 6 & number of covalent bonds [0,1,2,3,4,5] & Softmax & CE \textsuperscript{\emph{b}}\\
formal charge & integer & 1 & electrical charge & / & MSE \textsuperscript{\emph{c}} \\
radical electrons & integer & 1 & number of radical electrons & ReLU & MSE \\
hybridization & one-hot & 6 & [sp, sp$^2$, sp$^3$, sp$^3$d, sp$^3$d$^2$, other] & Softmax & CE \\
aromaticity & binary & 1 & whether the atom is part of an aromatic system [0/1] & Sigmoid & CE \\
hydrogens & one-hot & 5 & number of connected hydrogens [0,1,2,3,4] & Softmax & CE \\
chirality & binary & 1 & whether the atom is chiral center [0/1] & Sigmoid & CE \\
chirality type & binary & 2 & [R,S] & Sigmoid & CE \\
\hline
bond feature & type & size & description & $\phi_b$ & $L_b$\\
\hline
bond type & one-hot & 4 & [single, double, triple, aromatic] & Softmax & CE \\
conjugation & binary & 1 & whether the bond is conjugated [0/1] & Sigmoid & CE \\
ring & binary & 1 & whether the bond is in ring [0/1] & Sigmoid & CE \\
stereo & one-hot & 4 & [StereoNone, StereoAny, StereoZ, StereoE] & Softmax & CE \\
\hline
\end{tabular}}

\textsuperscript{\emph{a}} “WCE” means the weighted cross-entropy loss.
\textsuperscript{\emph{b}} “CE” means cross-entropy loss.
\textsuperscript{\emph{c}} “MSE” means mean square error.
\end{table*}

To discover the key atomic information in the molecule, Eq. \ref{eqAttention} perceives the relationship information $\mathbf{r}_i$ between the hidden layer vector $\mathbf{g}_i$ of each atom and the molecular embedding $\mathbf{f}$ through neural networks. Among them, $\mathbf{W}$ and $\mathbf{c}$ represent the weight matrix and bias vector of different neural networks (NNs); $[\cdot,\cdot]$ represents vector concatenation; “dropout” means to randomly drop some network nodes in the training batch to prevent overfitting; “elu” represents Exponential Linear Unit, which is used to activate the output of neurons nonlinearly, while retaining the nonlinear activation value of the negative part. \\
(4) Relation Readout:
\begin{equation}
\begin{aligned}
\mathbf{g}^{t-1}_i=\text{relu}(\text{GRU}(\mathbf{r}^t_i, \mathbf{g}^t_i)).
\label{eqReadout}
\end{aligned}
\end{equation}

To read out the embeddings $\mathbf{g}^{t-1}_i$ that are closer to the initial features of the atoms, Eq. \ref{eqReadout} is derived from the atomic current embeddings $\mathbf{g}^t_i$ and the relation information $\mathbf{r}^t_i$ through the Gated Recurrent Unit (GRU). “Relu” stands for Rectified Linear Unit, which is used for the output of non-linearly activated neurons. \\
(5) Attention:
\begin{equation}
\begin{aligned}
w_{N(i)}=\text{softmax}(\text{leaky\_relu}(\mathbf{W}\cdot\text{dropout}([\mathbf{g}^l_i, \mathbf{g}^l_{N(i)}])+\mathbf{c})).
\label{eqAttention}
\end{aligned}
\end{equation}

To focus on neighboring atoms that are critical to inferring initial features, Eq. \ref{eqAttention} gets the attention weight $w_{N(i)}$ between each atom $i$ and its adjacent atoms $N(i)$ by one-layer NNs. Among them, the initial values of $\mathbf{g}^l_i$ and $\mathbf{g}^l_{N(i)}$ come from $\mathbf{g}^{t=0}_i$ and $\mathbf{g}^{t=0}_{N(i)}$ in Eq. \ref{eqReadout}.\\
(6) Aggregation:
\begin{equation}
\begin{aligned}
\mathbf{C}^l_i=\text{elu}(\sum_{N(i)}w_{N(i)}\cdot\mathbf{W}(\text{dropout}(\mathbf{g}^l_{N(i)}))+\mathbf{c}).
\label{eqAggregation}
\end{aligned}
\end{equation}

To assist in inferring the initial features of atoms and bonds, Eq. \ref{eqAggregation} aggregates the context feature of each atom.\\
(7) Context Readout:
\begin{equation}
\begin{aligned}
\mathbf{g}^{l-1}_i=\text{relu}(\text{GRU}(\mathbf{C}^l_i, \mathbf{g}^l_i)).
\label{eqCReadout}
\end{aligned}
\end{equation}

To approximate the initial features, Eq. \ref{eqCReadout} deduces the hidden feature on each atom by GRU.\\
(8) Context Updating:
\begin{equation}
\begin{aligned}
\mathbf{g}^l_{N(i)}=\text{leaky\_relu}(\mathbf{W}\cdot\text{dropout}([\mathbf{g}^{l-1}_{i},\mathbf{g}^{l-1}_{N(i)}])+\mathbf{c}).
\label{eqCUpdating}
\end{aligned}
\end{equation}

For the next round of inference, Eq. \ref{eqCUpdating} updates the adjacent features for each atom by one-layer NNs.\\
(9) Generating:
\begin{equation}
\begin{aligned}
\hat{\mathbf{a}}_{i}=\phi_a(\mathbf{W}\cdot\mathbf{g}^0_{i}+\mathbf{c}),
\label{eqGenerateA}
\end{aligned}
\end{equation}
\begin{equation}
\begin{aligned}
\hat{\mathbf{b}}_{i,j}=\phi_b(\text{leaky\_relu}(\mathbf{W}\cdot\text{dropout}([\mathbf{g}^0_{i},\mathbf{g}^0_{j}])+\mathbf{c})).
\label{eqGenerateB}
\end{aligned}
\end{equation}

To generate the molecular graph, Eq. \ref{eqGenerateA} and \ref{eqGenerateB} predict the initial features of each atom and bond respectively through neural networks and an activation function. In Eq. \ref{eqGenerateA} and \ref{eqGenerateB}, $\phi_a$ and $\phi_b$ map the hidden embeddings to the initial features of atoms and bonds, respectively. The definitions of initial features of atoms and bonds are shown in table \ref{tInitialF}.

Denote AGRNs as a graph decoder $\mathrm{G}$, and its function to generate molecular graphs can be expressed as follows:
\begin{equation}
\begin{aligned}
&\mathrm{G}: \{\mathbf{f},\ \mathcal{H}\}\rightarrow\{\hat{\mathbf{a}}_{i},\ \mathcal{B}_i\}^{N_{a}}_{i=1},\\
&\text{where } \mathcal{H}=\{\mathbf{h}_i\}^{N_{a}}_{i=1},\ \mathcal{B}_i=\{\hat{\mathbf{b}}_{i,j}\}^{N(i)}_{j=1},
\label{eqAGRNs}
\end{aligned}
\end{equation}where $N_{a}$ represents the number of atoms contained in the molecule, and $N(i)$ represents the number of the adjacent atoms of the $i$-th atom.

\subsection{Generative Adversarial Feature Subspace Enhancement}
\label{secGAFSE}
\emph{1) Adversarial Feature Subspace Enhancement (AFSE) Algorithm: }AFSE was first proposed by us in \cite{yin2022afse} to enhance the model generalization for regression learning of molecular bioactivities. This paper extends AFSE to enhance model generalization for both regression and classification learning:
\begin{equation}
\begin{aligned}
\mathcal{L}_{\text{AFSE}}(\mathbf{f}, \mathrm{N}, \mathbf{d})&:=D(\mathrm{N}(\mathbf{f}, \mathbf{f}), \mathrm{N}(\mathbf{f}, \mathbf{f} \oplus \mathbf{d})), \\
\text { where } \mathbf{d}=\eta \frac{\mathbf{g}}{\|\mathbf{g}\|},\mathbf{g}&=\nabla_{\mathbf{r}} D(\mathrm{N}(\mathbf{f}, \mathbf{f}), \mathrm{N}(\mathbf{f}, \mathbf{f} \oplus \mathbf{r}))\rvert_{\|\mathrm{r}\| \leq \varepsilon},\\
D(\mathrm{N}(\mathbf{f}, \mathbf{f}), \mathrm{N}(\mathbf{f}, \mathbf{f} \oplus \mathbf{r}))&\triangleq[\sigma(\frac{\operatorname{N}(\mathbf{f},\mathbf{f}+\mathbf{r})+\varepsilon}{\operatorname{N}(\mathbf{f},\mathbf{f})+\varepsilon})-\gamma]^2\\
&+[\sigma(\frac{\operatorname{N}(\mathbf{f},\mathbf{f}-\mathbf{r})+\varepsilon}{\operatorname{N}(\mathbf{f},\mathbf{f})+\varepsilon})-\gamma]^2,\\
\operatorname{N}(\mathbf{f},\mathbf{f})&:=\text{Property},\\
\mathbf{f}&=\mathrm{E}(\mathbf{m}),
\label{eqAFSE}
\end{aligned}
\end{equation}where $\mathbf{m}$ is the initial graph representation of the molecule, $\mathbf{f}$ is the molecular embedding obtained by the graph learning model $\mathrm{E}$ according to $\mathbf{m}$, $\mathrm {N}$ is the neural network for downstream property regression or classification, $\mathbf{r}$ is a Gaussian random vector, $\eta$ is the learning rate, $\varepsilon$ is a small positive real number, $\sigma$ is the sigmoid function, $\gamma=\sigma(1)$ is used for the standardized discrepancy function $D(\cdot,\cdot)$. The adversarial perturbation $\mathbf{d}$ generated in Eq. \ref{eqAFSE} can significantly change the predicted property. After maximizing $\mathrm{N}$ and minimizing the AFSE loss for $\mathrm{E}$, $\mathrm{E}$ can extract more smooth embeddings for molecules with similar activity values but large structural differences to enhance generalization. At the same time, $\mathrm{N}$ can obtain the ability to perceive the effect of small changes in molecular embedding on the activity value. Therefore, the adversarial perturbation $\mathbf{d}$ obtained by Eq. \ref{eqAFSE} is an estimate of the potentially highly active molecular embedding $\mathbf{f}+\mathbf{d}$.

\emph{2) Molecular Graph Reconstruction: }Chemically, the fine-tuned molecules with significantly improved activity form matched molecular pairs with the original molecules, which are of great significance to the study of the activity cliff and the optimization of molecules. The most common chemical transformation is to modify a chemical element at an atomic site, perhaps the embedding $\mathbf{f}+\mathbf{d}$ of a potentially highly active molecule can lead us to intervene in a key atomic site for a replacement element. To this end, we use the AGRNs proposed in the previous section to reconstruct the molecule from $\mathbf{f}$; then find the key atomic positions and predict new chemical elements from $\mathbf{f}+\mathbf{d}$, which may optimize the property of the whole molecules. We design the reconstruction loss function to achieve this goal:
\begin{equation}
\begin{aligned}
\mathcal{L}_{Recon.}&(\hat{\mathbf{a}}_{i}, \mathbf{a}_{i}, \hat{\mathbf{b}}_{i,j}, \mathbf{b}_{i,j}):=\\
&\frac{1}{N_{a}}\sum^{N_{a}}_{i=1}\left[L_a(\hat{\mathbf{a}}_{i}, \mathbf{a}_{i})+\sum_{j\in N(i)} L_b(\hat{\mathbf{b}}_{i,j}, \mathbf{b}_{i,j})\right],
\label{eqLR}
\end{aligned}
\end{equation}where $L_a$ is the error function between reconstructed initial feature $\hat{\mathbf{a}}_{i}$, $\hat{\mathbf{b}}_{i,j}$ and the true initial feature $\mathbf{a}_{i}$ and $\mathbf{b}_{i,j}$ (see Table \ref{tInitialF}). The adjacent atom indexer $N(i)$ returns the atom index adjacent to the $i$-th atom.

It should be noted that, the content of various elements in natural molecules varies evidently, forming a priori distribution of chemical elements. Therefore, elements are weighted according to the proportion of atoms in each molecule when calculating the cross-entropy loss of node classification (denote as “WCE” in Table \ref{tInitialF} and used as $L_{a}$ in Eq. \ref{eqLR}). Specifically, the initial atom feature $\mathbf{a}_{i,k}=0/1$ can be expressed as whether the $i$-th atom belongs to the $k$-th element, and the reconstruction probability $\hat{\mathbf{a}}_{i,k}\in[0,1]$ measures the probability of the $i$-th atom belonging to the $k$-th element. Then their weighted cross-entropy loss can be defined as:
\begin{small}
\begin{equation}
\begin{aligned}
\operatorname{WCE}(\hat{\mathbf{a}}_{i,k}, \mathbf{a}_{i,k}):&=-\sum^{N_{a}}_{i=1}(1-\frac{N_{k^*(i)}}{N_{a}})\cdot\operatorname{CE}(\hat{\mathbf{a}}_{i}, \mathbf{a}_{i}),\\
\text{where } k^*(i)&=\mathop{\arg\max}_k \mathbf{a}_{i,k},
\label{eqWCE}
\end{aligned}
\end{equation}
\end{small}In Eq. \ref{eqWCE}, the initial element indexer $k^*(i)$ is a function of the atom index $i$. $N_{k^*}$ represents the number of the $k^*$-th element contained in all $N_{a}$ atoms. “CE” denotes the cross-entropy loss.

\emph{3) Molecular Graph Optimization: }
To provide stronger interpretability, the molecule optimization in this paper utilizes graph node classification logic to change the element symbols of one single atom at one time, forming matched molecular pairs on the activity cliff (MMP-Cliffs) \cite{cruz2014activity}. Therefore, the MMP-Cliffs generation scheme of GAFSE is illustrated in Figure \ref{framework}b. As shown, the binding sites of drug molecules to targets are often concentrated in a few key atomic sites, which usually have unique topological relationships relative to other atomic sites. Therefore, we first determine the key atomic positions through the distribution of posterior probabilities. Let $P(s\vert a)$ be the posterior probability that the atom $a$ is predicted to be the chemical element $s$, which can be estimated by the molecular embedding $\mathbf{f}$, the atomic embedding set $\mathcal{H}$, the graph decoder $\operatorname{G}:\mathbb{R}^{d_f}\times \mathbb{R}^{N_a\times d_f}\rightarrow [0,1]^{N_a\times N_s}$ and the $\mathbf{d}$ generated by the AFSE algorithm:
\begin{equation}
\begin{aligned}
P_{\mathbf{f}+\mathbf{d}}(s\vert a):&=\operatorname{G}(\mathbf{f}+\operatorname{Stopgrad}(\mathbf{d}),\mathcal{H})[a][s]=\widetilde{\mathbf{a}}_{i=a,k=s},\\
a&=1,2,\cdots,N_a,\ s=1,2,\cdots,N_s,
\label{eqPsa}
\end{aligned}
\end{equation}where “$\operatorname{Stopgrad}(\cdot)$” means to stop the propagation of the gradient, $N_a$ denotes the number of atoms, and $d_f$ denotes the dimension of the embedded features for both molecules and atoms. The number of element types $N_s$ is equal to 16, corresponding to the 1st to 16th dimension features of the atoms defined in Table \ref{tInitialF}. Then the key atomic position $a^*$ can be determined according to the maximum posterior probability criterion:
\begin{equation}
\begin{aligned}
&a^*=\mathop{\arg\max}_{a}\left(\max_{s\notin \mathcal{S}_a}P_{\mathbf{f}+\mathbf{d}}(s\vert a)\right),\\
&\text{where }\mathcal{S}_a:=\{s\vert P_{\mathbf{f}}(s\vert a)\geq P_0\}.
\label{eqa*}
\end{aligned}
\end{equation}In Eq. \ref{eqa*}, the conditional probability $P_{\mathbf{f}}(s\vert a)$ refers to the reconstruction probability $\hat{\mathbf{a}}_{i=a,k=s}$. Therefore, the set $\mathcal{S}_a$ represents the set of chemical elements at the $a$ atomic position predicted by graph generator $\operatorname{G}$ on the original molecular feature $\mathbf{f}$, whose probability is greater than the threshold $P_0$. According to our previous research \cite{yin2021metric, yin2021pseudo, yin2022universal}, after $\operatorname{G}$ is well trained, $\mathcal{S}_a$ usually contains initial elements and their confusing elements. Therefore, we hope to avoid the interference of these elements in the generation of new elements, i.e., let $s\notin \mathcal{S}_a$. Then, we determine the replaced element $s^*$ according to the maximum posterior probability criterion and the activation threshold $P_0$:
\begin{equation}
\begin{aligned}
s^*&=\mathop{\arg\max}_{s\in\mathcal{S}_a^*}P_{\mathbf{f}+\mathbf{d}}(s\vert a^*), \\
\text{where }\mathcal{S}_a^*:&=\{s\vert P_{\mathbf{f}+\mathbf{d}}(s\vert a^*)\geq P_0, s\neq s^0\}.
\label{eqs*}
\end{aligned}
\end{equation}In Eq. \ref{eqs*}, the candidate set $\mathcal{S}_a^*$ is the set of replaceable elements at atomic position $a^*$. Outside the candidate set $\mathcal{S}_a^*$, the correspond $a^*$ atom of $\max_s P_{\mathbf{f}+\mathbf{d}}(s\vert a^*)< P_0$ is not optimized to reduce the influence of low-confidence optimization on the result accuracy. Overall, a schematic diagram of GAFSE molecule optimization is shown (see Figure \ref{framework}c), and its implementation can be found in the supplementary Algorithm S3.

\emph{4) Validity Optimization of Generated Molecules: }The composition of molecules should conform to chemical specifications, so elements need to be verified for their validity when they are modified. To this end, this paper designs a novel mask-based validity optimization objective:
\begin{small}
\begin{equation}
\begin{aligned}
\mathcal{L}_{Val.}(a^*, s^*):=-\frac{1}{N_{a}}\sum^{N_{a}}_{i=1}(1-\operatorname{Val}(s_i^*\vert a_i^*))\cdot\log(1-P_{\mathbf{f}+\mathbf{d}}(s_i^*\vert a_i^*)).
\label{eqLV}
\end{aligned}
\end{equation}
\end{small}In Eq. \ref{eqLV}, the validity function $\operatorname{Val}(s_i^*\vert a_i^*)$ judges the chemical validity (1: valid, 0: invalid) of the optimized element $a_i^*$ (Eq. \ref{eqa*}) on selected atom $s_i^*$ (Eq. \ref{eqs*}) from the $i$-th molecule, and filter invalid molecules to calculate this validity loss. The probability $P_{\mathbf{f}+\mathbf{d}}(s_i^*\vert a_i^*)$ is defined by Eq. \ref{eqPsa}.

\emph{5) Generative Adversarial Feature Subspace Enhancement (GAFSE) Algorithm: }Finally, GAFSE presents four optimization objectives: (1) biological property objective $\mathcal{L}_{Bio.}$, (2) AFSE objective $\mathcal{L}_{AFSE}$, (3) reconstruction objective $\mathcal{L}_{Recon.}$ and (4) validity objective $\mathcal{L}_{Val.}$. In total, its optimization problem can be formulated as:
\begin{equation}
\begin{aligned}
&\min_{\operatorname{E},\operatorname{N},\operatorname{G}}\max_{\mathbf{d}}\ \underbrace{\mathcal{L}_{Bio.}+\lambda_1\mathcal{L}_{AFSE}}_{Representation\ Learning}+\lambda_2\underbrace{(\mathcal{L}_{Recon.}+\mathcal{L}_{Val.})}_{Molecular\ Optimization},\\
&\text{where }\mathcal{L}_{Bio.}:=\operatorname{Error}(\operatorname{N}(\mathbf{f},\mathbf{f}), \text{ Assay}),
\label{eqGAFSE}
\end{aligned}
\end{equation}where “Assay” represents the molecular activity or property value determined by chemical wet experiments; $\operatorname{Error}(\cdot, \text{Assay})$ represents the error function between predictions and assays, that is, the cross-entropy loss for classification assays or the mean square error for regression assays; The coefficient $\lambda_1$ balances the biological property loss and the AFSE loss, while $\lambda_2$ balances the representation learning and the molecule optimization.

\begin{figure*}[t]
\centering
\includegraphics[width=1\linewidth,height=3.5in]{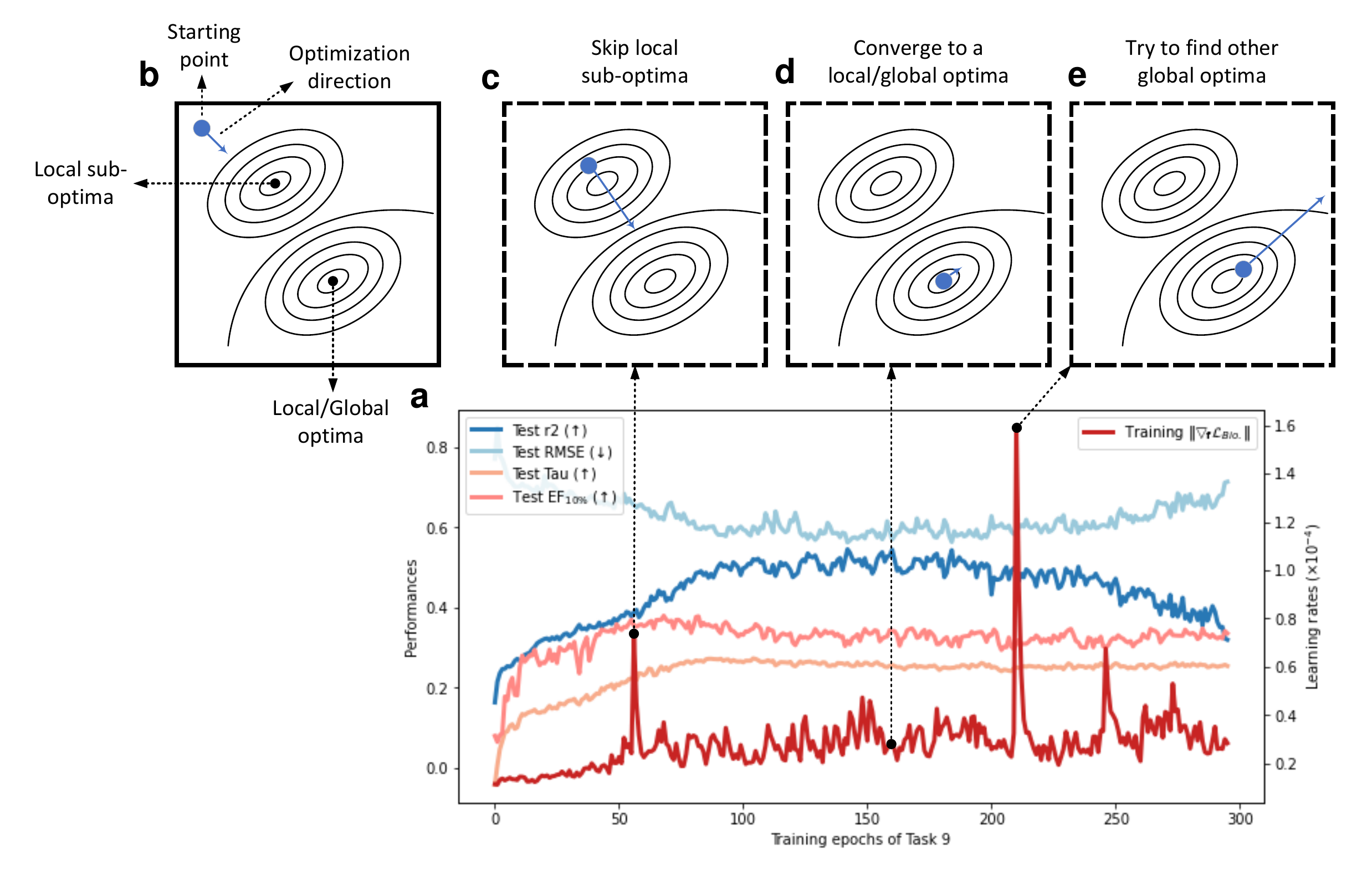}
\caption{\textbf{a}. The regulation and control of GAFSE-HS learning rate penalty factor $\Vert\nabla_{\mathbf{f}}\mathcal{L}_{Bio.}\Vert$ for model optimization. “($\uparrow$)” means larger is better and vice versa. \textbf{b}. Legend for model optimization diagram. \textbf{c}. Larger $\Vert\nabla_{\mathbf{f}}\mathcal{L}_{Bio.}\Vert$ forces the model to skip local sub-optima. \textbf{d}. A smaller $\Vert\nabla_{\mathbf{f}}\mathcal{L}_{Bio.}\Vert$ allows the model to converge to a local or possibly global optima. \textbf{e}. Larger $\Vert\nabla_{\mathbf{f}}\mathcal{L}_{Bio.}\Vert$ encourages the model to find other potential global optima.}
\label{punishlr}
\end{figure*}

\subsection{Convergence guarantee on representation learning}
\label{secConvergence}
To theoretically analyze the representation learning of the GAFSE framework, we performed a convergence analysis on it. Firstly, one layer of Neural Networks (NNs) with multiple outputs (or one output) can be composed of a weight matrix $\mathbf{W}$ (or vector $\mathbf{a}$) and a bias vector $\mathbf{c}$ (or variable $c$). For a clearer analysis, we omit the writing of the bias $\mathbf{c}$ and the constant $\gamma$ (Eq. \ref{eqAFSE}) in the following derivation. Note that the weight matrix $\mathbf{W}$ (or vector $\mathbf{a}$) is irreducible after omitting the bias $\mathbf{c}$. Two dimensionality reduction matrices $\mathbf{W}_1$ and $\mathbf{W_2}$ are introduced to map the two features $\mathbf{f}$ and $\mathbf{f}+\mathbf{d}$ to half of the input dimension of $\mathbf{a}$, respectively. Suppose the feature vector extracted by $\operatorname{Network}$ is $\mathbf{f}$, the biological property assay is $y$, and the object of representation learning can be written as
\begin{small}
\begin{equation}
\begin{aligned}
&\min_{\mathbf{a},\mathbf{W}_1,\mathbf{W}_2} \underbrace{(\mathbf{a[W_1,W_2]f}-y)^2}_{\mathcal{L}_{Bio.}}\\
&+\underbrace{\sigma(\frac{\mathbf{a[W_1f,W_2(f+\mathbf{d})]}}{\mathbf{a[W_1,W_2]f}})+\sigma(\frac{\mathbf{a[W_1f,W_2(f-\mathbf{d})]}}{\mathbf{a[W_1,W_2]f}})}_{\mathcal{L}_{\operatorname{AFSE}}}\\
&\text{where }\sigma(x)=(\frac{1}{1+e^{-x}}-1)^2=(\frac{e^{-x}}{1+e^{-x}})^2\\
\mathbf{d}&=\nabla_{\mathbf{r}}\ \sigma(\frac{\mathbf{a[W_1f,W_2(f+r)]}}{\mathbf{a[W_1,W_2]f}})+\sigma(\frac{\mathbf{a[W_1f,W_2(f-r)]}}{\mathbf{a[W_1,W_2]f}})\\
&\mathbf{r}\sim N(\mathbf{0},\mathbf{1}),
\label{opVSAT}
\end{aligned}
\end{equation}
\end{small}where $[\cdot,\cdot]$ means vector concatenation. The activation function $\sigma(x)$ we defined in the above formula is differentiable, and its derivative can be expressed by itself as
\begin{equation}
\begin{aligned}
\sigma^{\prime}(x)=\frac{2{\mathrm{e}}^{-2x}}{{\left(1+{\mathrm{e}}^{-x}\right)}^2}(\frac{{\mathrm{e}}^{-x}}{{{1+\mathrm{e}}^{-x}}}-1)=2\sigma(x)(\sqrt{\sigma(x)}-1).
\label{sigma'}
\end{aligned}
\end{equation}According to Eq. \ref{opVSAT}, the gradient $\mathbf{d}$ can be simplified to
\begin{small}
\begin{equation}
\begin{aligned}
\mathbf{d}=\frac{\mathbf{aW_2}}{\mathbf{a[W_1,W_2]f}}(&\sigma^{\prime}(\frac{\mathbf{a[W_1f,W_2(f+r)]}}{\mathbf{a[W_1,W_2]f}})\\
-&\sigma^{\prime}(\frac{\mathbf{a(W_1f+W_2(f-r))}}{\mathbf{a[W_1,W_2]f}})).
\label{nabla_d}
\end{aligned}
\end{equation}
\end{small}According to Eq. \ref{opVSAT}, there is $0<\sigma(x)<1$ for any $x$. Then according to Eq. \ref{sigma'}, it has $-2<\sigma^{\prime}(x)<0$ for any $x$. Therefore, the gradient $\mathbf{d}$ satisfies
\begin{equation}
\begin{aligned}
\frac{-2\mathbf{aW_2}}{\mathbf{a[W_1,W_2]f}}<\mathbf{d}<\frac{2\mathbf{aW_2}}{\mathbf{a[W_1,W_2]f}}.
\label{nabla_range}
\end{aligned}
\end{equation}After the gradient $\mathbf{d}$ is calculated, the gradient generated from $\mathbf{f}$ is cleared. Therefore, when the gradient of $\mathbf{f}$ for Eq. \ref{opVSAT} is calculated, $\mathbf{d}$ is regarded as a constant. Then, we have
\begin{equation}
\begin{aligned}
\nabla_{\mathbf{f}}\mathcal{L}_{\operatorname{AFSE}}=-\frac{\mathbf{aW_2}\nabla}{\mathbf{f}^2}(&\sigma^{\prime}(\frac{\mathbf{a[W_1f,W_2(f+\mathbf{d})]}}{\mathbf{a[W_1,W_2]f}})\\
+&\sigma^{\prime}(\frac{\mathbf{a[W_1f,W_2(f-\mathbf{d})]}}{\mathbf{a[W_1,W_2]f}})).
\label{LVSAT'}
\end{aligned}
\end{equation}Similarly, according to Eq. \ref{nabla_range}, $\nabla_{\mathbf{f}}\mathcal{L}_{\operatorname{AFSE}}$ satisfies
\begin{equation}
\begin{aligned}
\frac{-8\mathbf{a^2W_2^2}}{\mathbf{a[W_1,W_2]f^3}}<\nabla_{\mathbf{f}}\mathcal{L}_{\operatorname{AFSE}}<\frac{8\mathbf{a^2W_2^2}}{\mathbf{a[W_1,W_2]f^3}}.
\label{LVSAT_range}
\end{aligned}
\end{equation}For any two features $\mathbf{f_1}$ and $\mathbf{f_2}$, we have
\begin{small}
\begin{equation}
\begin{aligned}
\Vert\nabla_{\mathbf{f_1}}\mathcal{L}_{\operatorname{AFSE}}-\nabla_{\mathbf{f_2}}\mathcal{L}_{\operatorname{AFSE}}\Vert<\Vert\frac{8\mathbf{a^2W_2^2}}{\mathbf{a[W_1,W_2]}}(\frac{1}{\mathbf{f}_1^3}-\frac{1}{\mathbf{f}_2^3})\Vert&\\
=\Vert\frac{8\mathbf{a^2W_2^2}(\mathbf{f}_1^2+\mathbf{f}_1\mathbf{f}_2+\mathbf{f}_2^2)}{\mathbf{a[W_1,W_2]\mathbf{f}_1^3\mathbf{f}_2^3}}(\mathbf{f}_1-\mathbf{f}_2)\Vert.&
\label{convergence1}
\end{aligned}
\end{equation}
\end{small}Features are normalized, so it has $\Vert\mathbf{f}_1\Vert=\Vert\mathbf{f}_2\Vert=1$. According to the triangle inequality, Eq. \ref{convergence1} can be further simplified to
\begin{small}
\begin{equation}
\begin{aligned}
\Vert\nabla_{\mathbf{f_1}}\mathcal{L}_{\operatorname{AFSE}}-\nabla_{\mathbf{f_2}}\mathcal{L}_{\operatorname{AFSE}}\Vert&<\Vert\frac{24\mathbf{a^2W_2^2}}{\mathbf{a[W_1,W_2]}}\Vert\cdot\Vert\mathbf{f}_1-\mathbf{f}_2\Vert\\
&\triangleq\beta_{\operatorname{AFSE}}\Vert\mathbf{f}_1-\mathbf{f}_2\Vert.
\label{convergence2}
\end{aligned}
\end{equation}
\end{small}Similarily, for $\mathcal{L}_{Bio.}$, there is
\begin{small}
\begin{equation}
\begin{aligned}
\Vert\nabla_{\mathbf{f_1}}\mathcal{L}_{Bio.}-\nabla_{\mathbf{f_2}}\mathcal{L}_{Bio.}\Vert&<\Vert2\mathbf{a^2[W_1,W_2]^2}\Vert\cdot\Vert\mathbf{f}_1-\mathbf{f}_2\Vert\\
&\triangleq\beta_{\operatorname{Bio.}}\Vert\mathbf{f}_1-\mathbf{f}_2\Vert.
\label{convergence3}
\end{aligned}
\end{equation}
\end{small}Here, $\nabla_{\mathbf{f}}\mathcal{L}_{AFSE}$ and $\nabla_{\mathbf{f}}\mathcal{L}_{Bio.}$ are functions of $\mathbf{f}$. According to the triangle inequality \cite{khamsi2001introduction}, $\mathcal{L}_{Bio.}+\mathcal{L}_{AFSE}$ is \textit{Lipschitz continous} \cite{eriksson2004lipschitz} on $\mathcal{F}=\{\mathbf{f}\in\mathbb{R}^{d_f}:\Vert\mathbf{f}\Vert=1\}$ with the \textit{Lipschitz constent} of $\beta_{\operatorname{AFSE}}+\beta_{\operatorname{Bio.}}$. That is, for all $\mathbf{f_1},\mathbf{f_2}\in\mathcal{F}$, it has
\begin{small}
\begin{equation}
\begin{aligned}
&\Vert\nabla_{\mathbf{f_1}}(\mathcal{L}_{\operatorname{AFSE}}+\mathcal{L}_{Bio.})-\nabla_{\mathbf{f_2}}(\mathcal{L}_{\operatorname{AFSE}}+\mathcal{L}_{Bio.})\Vert\\
&\leq\Vert\nabla_{\mathbf{f_1}}\mathcal{L}_{\operatorname{AFSE}}-\nabla_{\mathbf{f_2}}\mathcal{L}_{\operatorname{AFSE}}\Vert+\Vert\nabla_{\mathbf{f_1}}\mathcal{L}_{Bio.}-\nabla_{\mathbf{f_2}}\mathcal{L}_{Bio.}\Vert\\
&<(\beta_{\operatorname{AFSE}}+\beta_{\operatorname{Bio.}})\Vert\mathbf{f}_1-\mathbf{f}_2\Vert.
\label{eqLipschitz}
\end{aligned}
\end{equation}
\end{small}Let $\mathbf{f}_{t}$ be the learnable feature at $t$-th optimization step. Then $\mathbf{f}_{t+1}$ can be updated by
\begin{equation}
\begin{aligned}
\mathbf{f}_{t+1}\leftarrow\mathbf{f}_{t}-\eta\nabla_{\mathbf{f_{t}}}(\mathcal{L}_{\operatorname{AFSE}}+\mathcal{L}_{Bio.}),
\label{equpdate}
\end{aligned}
\end{equation}where $\eta$ is the learning rate. According to Eq. \ref{eqLipschitz}, it has
\begin{small}
\begin{equation}
\begin{aligned}
&\Vert\nabla_{\mathbf{f_{t+1}}}(\mathcal{L}_{\operatorname{AFSE}}+\mathcal{L}_{Bio.})-\nabla_{\mathbf{f_t}}(\mathcal{L}_{\operatorname{AFSE}}+\mathcal{L}_{Bio.})\Vert\\
&=\frac{1}{\eta}\Vert\eta\nabla_{\mathbf{f_{t+1}}}(\mathcal{L}_{\operatorname{AFSE}}+\mathcal{L}_{Bio.})-\eta\nabla_{\mathbf{f_t}}(\mathcal{L}_{\operatorname{AFSE}}+\mathcal{L}_{Bio.})\Vert\\
&=\frac{1}{\eta}\Vert\mathbf{f_{t+2}}-\mathbf{f_{t+1}}+\mathbf{f_{t+1}}-\mathbf{f_{t}}\Vert\\
&=\frac{1}{\eta}\Vert\mathbf{f_{t+2}}-\mathbf{f_{t}}\Vert\\
&<(\beta_{\operatorname{AFSE}}+\beta_{\operatorname{Bio.}})\Vert\mathbf{f}_{t+1}-\mathbf{f}_t\Vert.
\label{eqft}
\end{aligned}
\end{equation}
\end{small}Then,
\begin{equation}
\begin{aligned}
\Vert\mathbf{f_{t+2}}-\mathbf{f_{t}}\Vert<\eta(\beta_{\operatorname{AFSE}}+\beta_{\operatorname{Bio.}})\Vert\mathbf{f}_{t+1}-\mathbf{f}_t\Vert.
\label{eqft2}
\end{aligned}
\end{equation}Suppose $\mathbf{f}^*$ be the global optimal solution. If $\mathbf{f_{t}}=\mathbf{f}^*$, when $\eta<\frac{1}{\beta_{\operatorname{AFSE}}+\beta_{\operatorname{Bio.}}}$, Eq. \ref{opVSAT} has convergence:
\begin{equation}
\begin{aligned}
\left\|\mathbf{f}_{t+2}-\mathbf{f}^{*}\right\|^{2}<\left\|\mathbf{f}_{t+1}-\mathbf{f}^{*}\right\|^{2}.
\label{convergence4}
\end{aligned}
\end{equation}If $\mathbf{f_{t}}\neq\mathbf{f}^*$, a larger learning rate, $\eta>\frac{1}{\beta_{\operatorname{AFSE}}+\beta_{\operatorname{Bio.}}}$, should be employed to encourage $\mathbf{f_{t+1}}$ stay away from $\mathbf{f_{t}}$:
\begin{equation}
\begin{aligned}
\left\|\mathbf{f}_{t+2}-\mathbf{f}_t\right\|^{2}<L_f\left\|\mathbf{f}_{t+1}-\mathbf{f}_t\right\|^{2} \text{, where $L_f>1$}.
\label{convergence5}
\end{aligned}
\end{equation}To this end, we define the learning rate $\eta$ of $\mathbf{a}$, $\mathbf{W_1}$ and $\mathbf{W_2}$ as:
\begin{small}
\begin{equation}
\begin{aligned}
\eta=\left\{
\begin{array}{lr}
\eta^*+\alpha\Vert\nabla_{\mathbf{f}}\mathcal{L}_{Bio.}\Vert-o(\eta^*), & \eta^*<\eta_{max} \\
\eta_{max}, & \eta^*\geq\eta_{max}
\end{array}
\right.,
\label{eqeta}
\end{aligned}
\end{equation}
\end{small}where $\eta^*=\frac{1}{\beta_{\operatorname{AFSE}}+\beta_{\operatorname{Bio.}}}$. $\eta_{max}$ is the maximum learning rate used to stabilize the model parameters. $\alpha$ is the balance coefficient. When the gradient of biological property loss decays to $o(\eta^*)/\alpha$, it can be considered that $\mathbf{f_{t}}$ approximates $\mathbf{f}^*$. Then, we have $\eta\leq\eta^*$ according to Eq. \ref{eqeta}, that is, Eq. \ref{opVSAT} converges.

When calculating Eq. \ref{eqeta}, the learning rate $\eta$ needs to be determined before the backward propagation to update the network parameters, so we need to calculate $\beta_{\operatorname{AFSE}}$ and $\beta_{\operatorname{Bio.}}$ in the forward propagation of the neural network to ensure the convergence of GAFSE. To this end, we use the unit vector $\mathbf{1}$ and the zero vector $\mathbf{0}$ to derive the operation of the neural network parameters, namely $\beta_{\operatorname{AFSE}}$ and $\beta_{ \operatorname{Bio.}}$ can be calculated as:
\begin{equation}
\begin{aligned}
\beta_{\operatorname{AFSE}}=\Vert\frac{24\cdot(\mathbf{a}^T\times \mathbf{[W_1,W_2]\times [0,1]})^2}{\mathbf{a}^T\times\mathbf{[W_1,W_2]\times [1,1]}}\Vert^2,
\label{eqbetaAFSE}
\end{aligned}
\end{equation}
\begin{equation}
\begin{aligned}
\beta_{\operatorname{Bio.}}=4\cdot\Vert\mathbf{a}^T\times\mathbf{[W_1,W_2]\times [1,1]}\Vert^4.
\label{eqbetaMSE}
\end{aligned}
\end{equation}In each step of updating the neural network parameters by statistical gradient descent on $\mathcal{L}_{Bio.}$ and $\mathcal{L}_{AFSE}$, Eq. \ref{eqeta}, \ref{eqbetaAFSE} and \ref{eqbetaMSE} determine the learning rate $\eta$ to ensure the convergence of the representation learning on $\mathbf{f}$.

\begin{figure*}[t]
\centering
\includegraphics[width=1\linewidth,height=3.5in]{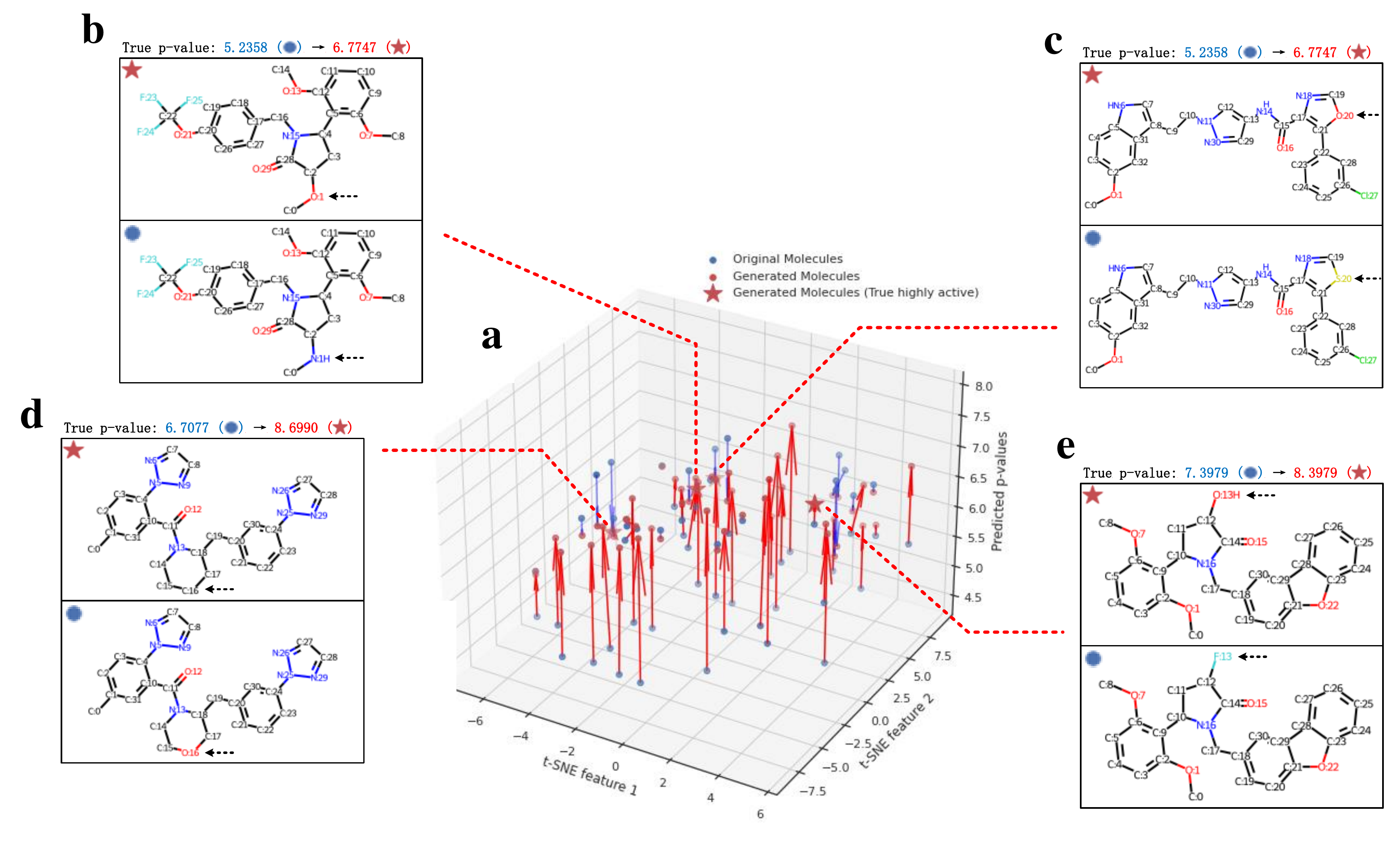}
\caption{Location of GAFSE-generated molecules in the predicted structural feature-activity space for binding to the orexins receptor O43614 from the \href{http://www.uniprot.org/docs/7tmrlist}{UniProt} database. The red arrow indicates that the predicted activity value of the generated molecule is higher than that of their original molecules, and the blue arrow is the opposite. \textbf{b-e}. The generated true MMP-Cliffs in the dataset.}
\label{t-sne}
\end{figure*}

\newcommand{\tabincell}[2]{\begin{tabular}{@{}#1@{}}#2\end{tabular}}
\begin{table*}[!h]
\centering
\caption{Comparison of ligand-based virtual screening indexes on benchmark datasets \cite{yin2022afse}. Baseline results are taken from \cite{yin2022afse}.}
\resizebox{\textwidth}{50mm}{
\begin{tabular}{cccccccccccccc}
\hline
\multirow{2}{*}{Dataset size} & \multicolumn{1}{l}{\multirow{2}{*}{\tabincell{c}{Task ID\\(Bias $\uparrow$)}}} & \multicolumn{3}{c}{EF$_{10\%}$ (\%) ($\uparrow$)} & \multicolumn{3}{c}{$r^2$ ($\uparrow$)} & \multicolumn{3}{c}{RMSE ($\downarrow$)} & \multicolumn{3}{c}{$\tau_B$ ($\uparrow$)} \\
\cmidrule(lr){3-5}\cmidrule(lr){6-8}\cmidrule(lr){9-11}\cmidrule(lr){12-14}& & AFSE & GAFSE-HS & GAFSE-MO & \multicolumn{1}{c}{AFSE} & \multicolumn{1}{c}{GAFSE-HS} & GAFSE-MO & AFSE & GAFSE-HS & GAFSE-MO & AFSE & GAFSE-HS & GAFSE-MO \\
\hline
\multirow{7}[2]{*}{small} & 1 & 38.46 & 26.92 & 26.92 & \multicolumn{1}{c}{0.4347 } & \multicolumn{1}{c}{0.4564 } & 0.4621 & 0.6619 & 0.6112 & 0.6058 & 0.0772 & 0.0508 & 0.0491 \\
& 2 & 27.78 & 44.44 & 44.44 & \multicolumn{1}{c}{0.1030 } & \multicolumn{1}{c}{0.1187 } & 0.1485 & 0.6618 & 0.9844 & 0.8089 & 0.3794 & 0.4624 & 0.4480 \\
& 3 & 57.89 & 52.63 & 63.16 & \multicolumn{1}{c}{0.2774 } & \multicolumn{1}{c}{0.4106 } & 0.4399 & 1.0085 & 0.8595 & 0.8347 & 0.2671 & 0.4310 & 0.4437 \\
& 4 & 62.50 & 65.62 & 62.50 & \multicolumn{1}{c}{0.4660 } & \multicolumn{1}{c}{0.4648 } & 0.4508 & 0.7126 & 0.6854 & 0.6980 & 0.3986 & 0.3898 & 0.3952 \\
& 5 & 53.85 & 53.85 & 46.15 & \multicolumn{1}{c}{0.6423 } & \multicolumn{1}{c}{0.6241 } & 0.6256 & 0.7391 & 0.8315 & 0.7595 & 0.4222 & 0.3883 & 0.3774 \\
& 6 & 38.46 & 53.85 & 61.54 & \multicolumn{1}{c}{0.2022 } & \multicolumn{1}{c}{0.4563 } & 0.4490 & 0.8763 & 0.9631 & 0.9306 & 0.4116 & 0.4435 & 0.4513 \\
& 7 & 64.29 & 71.43 & 71.43 & \multicolumn{1}{c}{0.5079 } & \multicolumn{1}{c}{0.5898 } & 0.5842 & 0.8799 & 0.8081 & 0.7754 & 0.5293 & 0.5085 & 0.5249 \\
\hline
\multirow{10}[2]{*}{Median} & 8 & 68.24 & 69.41 & 67.06 & \multicolumn{1}{c}{0.6201 } & \multicolumn{1}{c}{0.6050 } & 0.6068 & 0.8900 & 0.9150 & 0.9066 & 0.5783 & 0.5711 & 0.5842 \\
& 9 & 37.00 & 34.00 & 35.00 & \multicolumn{1}{c}{0.5018 } & \multicolumn{1}{c}{0.5177 } & 0.5292 & 0.5764 & 0.5675 & 0.5623 & 0.2823 & 0.2597 & 0.2668 \\
& 10 & 44.23 & 50.00 & 51.92 & \multicolumn{1}{c}{0.5102 } & \multicolumn{1}{c}{0.5158 } & 0.4980 & 0.7379 & 0.7104 & 0.7212 & 0.2915 & 0.3680 & 0.3546 \\
& 11 & 47.54 & 47.54 & 49.18 & \multicolumn{1}{c}{0.3669 } & \multicolumn{1}{c}{0.3445 } & 0.3979 & 0.9792 & 0.9381 & 0.8959 & 0.3113 & 0.3043 & 0.3307 \\
& 12 & 52.87 & 51.72 & 56.32 & \multicolumn{1}{c}{0.5926 } & \multicolumn{1}{c}{0.5954 } & 0.5887 & 0.9182 & 0.9353 & 0.9469 & 0.4318 & 0.4414 & 0.4410 \\
& 13 & 46.94 & 52.04 & 51.02 & \multicolumn{1}{c}{0.3702 } & \multicolumn{1}{c}{0.3285 } & 0.4020 & 0.8425 & 0.8932 & 0.8628 & 0.4641 & 0.4628 & 0.4569 \\
& 14 & 74.32 & 71.62 & 75.68 & \multicolumn{1}{c}{0.6144 } & \multicolumn{1}{c}{0.5936 } & 0.6057 & 0.9802 & 1.0619 & 1.0112 & 0.3908 & 0.4092 & 0.4203 \\
& 15 & 40.74 & 37.04 & 37.04 & \multicolumn{1}{c}{0.3943 } & \multicolumn{1}{c}{0.4369 } & 0.3895 & 0.8969 & 0.8316 & 0.8640 & 0.2056 & 0.2031 & 0.1679 \\
& 16 & 69.39 & 66.33 & 65.31 & \multicolumn{1}{c}{0.6416 } & \multicolumn{1}{c}{0.5926 } & 0.5870 & 0.7186 & 0.8366 & 0.8331 & 0.4410 & 0.4051 & 0.4001 \\
& 17 & 62.92 & 51.69 & 59.55 & \multicolumn{1}{c}{0.3032 } & \multicolumn{1}{c}{0.2124 } & 0.2849 & 0.9487 & 0.9273 & 0.9295 & 0.5322 & 0.5150 & 0.5338 \\
\hline \multirow{17}[4]{*}{Large} & 18 & 64.94 & 55.84 & 58.44 & \multicolumn{1}{c}{0.5766 } & \multicolumn{1}{c}{0.5473 } & 0.5682 & 0.8008 & 0.8389 & 0.8315 & 0.4414 & 0.4153 & 0.4299 \\
& 19 & 63.41 & 65.85 & 65.85 & \multicolumn{1}{c}{0.5188 } & \multicolumn{1}{c}{0.5318 } & 0.5318 & 0.6612 & 0.6503 & 0.6503 & 0.6701 & 0.6677 & 0.6677 \\
& 20 & 65.85 & 65.04 & 65.04 & \multicolumn{1}{c}{0.5152 } & \multicolumn{1}{c}{0.6504 } & 0.5664 & 0.6918 & 0.7088 & 0.6546 & 0.4422 & 0.4507 & 0.4694 \\
& 21 & 61.34 & 57.14 & 56.30 & \multicolumn{1}{c}{0.5173 } & \multicolumn{1}{c}{0.5284 } & 0.5195 & 0.7244 & 0.7071 & 0.7201 & 0.4716 & 0.7071 & 0.4583 \\
& 22 & 57.35 & 63.24 & 63.24 & \multicolumn{1}{c}{0.5291 } & \multicolumn{1}{c}{0.5343 } & 0.5062 & 0.7201 & 0.7299 & 0.7553 & 0.4071 & 0.4098 & 0.4048 \\
& 23 & 58.02 & 61.83 & 62.60 & \multicolumn{1}{c}{0.5674 } & \multicolumn{1}{c}{0.5661 } & 0.5645 & 0.7491 & 0.7579 & 0.7560 & 0.4833 & 0.4899 & 0.4919 \\
& 24 & 62.59 & 64.03 & 69.06 & \multicolumn{1}{c}{0.6161 } & \multicolumn{1}{c}{0.5994 } & 0.6288 & 0.7799 & 0.8024 & 0.7870 & 0.5676 & 0.5434 & 0.5666 \\
& 25 & 68.66 & 63.43 & 66.42 & \multicolumn{1}{c}{0.7110 } & \multicolumn{1}{c}{0.5700 } & 0.5779 & 0.5991 & 0.7451 & 0.7364 & 0.4271 & 0.3970 & 0.4220 \\
& 26 & 67.61 & 69.01 & 67.61 & \multicolumn{1}{c}{0.5809 } & \multicolumn{1}{c}{0.5838 } & 0.5371 & 0.7079 & 0.7528 & 0.7797 & 0.5061 & 0.5301 & 0.5370 \\
& 27 & 60.75 & 58.88 & 59.81 & \multicolumn{1}{c}{0.5522 } & \multicolumn{1}{c}{0.5684 } & 0.5484 & 0.8023 & 0.7006 & 0.7127 & 0.5230 & 0.5108 & 0.4972 \\
& 28 & 50.00 & 57.50 & 53.75 & \multicolumn{1}{c}{0.4087 } & \multicolumn{1}{c}{0.4901 } & 0.4714 & 0.7009 & 0.6441 & 0.6811 & 0.3758 & 0.6441 & 0.4100 \\
& 29 & 63.70 & 62.22 & 57.04 & \multicolumn{1}{c}{0.6177 } & \multicolumn{1}{c}{0.6222 } & 0.5415 & 0.7059 & 0.7884 & 0.7768 & 0.3253 & 0.3077 & 0.2908 \\
& 30 & 43.75 & 60.71 & 41.07 & \multicolumn{1}{c}{0.3941 } & \multicolumn{1}{c}{0.6071 } & 0.4086 & 0.9048 & 0.8987 & 0.8923 & 0.2898 & 0.3766 & 0.2390 \\
& 31 & 53.01 & 51.20 & 48.19 & \multicolumn{1}{c}{0.4972 } & \multicolumn{1}{c}{0.4112 } & 0.4190 & 0.7193 & 0.7622 & 0.7583 & 0.3337 & 0.3064 & 0.3005 \\
& 32 & 61.54 & 61.54 & 62.82 & \multicolumn{1}{c}{0.5883 } & \multicolumn{1}{c}{0.6665 } & 0.6148 & 0.8863 & 0.6665 & 0.8616 & 0.3519 & 0.3851 & 0.3702 \\
& 33 & 58.55 & 54.40 & 56.48 & \multicolumn{1}{c}{0.5025 } & \multicolumn{1}{c}{0.5440 } & 0.5614 & 0.6984 & 0.6594 & 0.6840 & 0.3076 & 0.6594 & 0.3214 \\
\cline{2-14} & Average & 56.01 & 56.73 & \textbf{56.91 } & \multicolumn{1}{c}{0.4922} & \multicolumn{1}{c}{\textbf{0.5116}} & 0.5035 & \textbf{0.7843 } & 0.7931 & \textbf{0.7874} & 0.4042 & \textbf{0.4368 } & 0.4098 \\
\hline
\end{tabular}}
\label{tPerformance}
\end{table*}

\section{Experiments}
\subsection{Screening of highly active molecules}
\label{secScreening}
\emph{Setup: } To validate the ability of GAFSE to screen highly active molecules in compound databases, we evaluate the virtual screening performance of GAFSE on the benchmark dataset constructed in \cite{yin2022afse}. Meanwhile, to validate whether the convergence learning rate $\eta$ in Eq. \ref{eqeta} encourages the AFSE method to converge to a better solution, we only take the “representation learning” part of Eq. \ref{eqGAFSE}, degenerating it an improved method of AFSE (GAFSE-HS), and compare their performance in this section. Furthermore, we found that the model works stably when the balance coefficients in Eq. \ref{eqGAFSE} and Eq. \ref{eqeta} are taken as 0.6, 0.3, and 0.06. Therefore, we fixed $\lambda_1=0.6$, $\lambda_2=0.3$ and $\alpha=0.06$ in all subsequent experiments. To reduce the computational cost, experiments in this paper omit the reconstruction loss except for atomic symbols, which are optimized for molecule optimization.

\emph{Algorithm Review: }This section is about GAFSE experiments in the hit screening (HS) stage of lead discovery, referred to as GAFSE-HS. The pipeline of GAFSE-HS is as follows: First, molecules are input into the molecular embedding model $\operatorname{E}$ in the form of graphs to obtain molecular embeddings $\mathbf{f}$; then $\mathbf{f}$ is input to the downstream task model $\operatorname{N}$ to output predicted bioactivity values $\hat{y}$. In the training phase of GAFSE-HS, the output $\hat{y}$ participates in the optimization of the “representation learning” part of Eq. \ref{eqGAFSE}, thereby updating model parameters of $\operatorname{E}$ and $\operatorname{N}$. In the testing phase of GAFSE-HS, the output $\hat{y}$ is used to screen out highly active molecules to complete the HS task. In this section, the performance of GAFSE-MO (introduced in Section \ref{secAO}: Algorithm Review) on the HS task is additionally carried out to explore the impact of molecule optimization on representation learning.

\emph{Results:} Taking the lysolipids receptor Q99500 in the \href{http://www.uniprot.org/docs/7tmrlist}{UniProt} database as an example, Figure \ref{punishlr} shows the change in GAFSE-HS learning rate penalty factor $\Vert\nabla_{\mathbf{f}}\mathcal{L}_{Bio.}\Vert$ (see Eq. \ref{eqeta}) and the corresponding test performance. The results show that the training penalty factor $\Vert\nabla_{\mathbf{f}}\mathcal{L}_{Bio.}\Vert$ can be reduced when the potential test performance improves, so that the model can be trained at the convergent learning rate $\gamma^*$ (see Eq. \ref{eqeta}); and increases when the potential test performance decreases to jump out of the local optima and find the potential global optima; finally stops the model optimization after exploring a certain number of steps, and the converged model parameters are taken, which effectively enhances the generalization of the model.

To verify that GAFSE-HS and GAFSE-MO can achieve better generalization performance than AFSE \cite{yin2022afse}, we compare four ligand-based virtual screening indexes of each method on benchmark datasets. As shown in Table \ref{tPerformance}, when GAFSE-HS is compared with AFSE, the index (EF$_{10\%}$) of screening highly active molecules (EF$_{10\%}$) on all 33 tasks improved by an average of 1.29\%; the index ($r^2$) for fitting the distribution of activity values increased by an average of 3.94\%; the index ($\tau_B$) for predicting the ranking of activity values increased by an average of 8.07\%. These results show that the theoretically guaranteed learning rate definition (Eq. \ref{eqeta}) of GAFSE-HS in this paper can generally improve the generalization performance of AFSE. Compared with GAFSE-HS, the average EF$_{10\%}$ and RMSE of GAFSE-MO are further improved by 0.32\% and 0.73\% on all 33 tasks, respectively. These results show that the reconstruction of the molecular graph by GAFSE-MO does not negatively affect the prediction of absolute activity values (RMSE) and the hit rate of highly active molecules (EF$_{10\%}$).

\begin{table*}
\centering
\caption{Generation of higher active drug molecules by GAFSE-MO}
\resizebox{\textwidth}{90mm}{
\begin{tabular}{ccccc}
\hline
Targets & Anchors & Anchor Properties & Generated molecules & Generated Properties\\
\hline
\multirow{21}{*}{\tabincell{c}{Human\\sphingosine\\1-phosphate\\receptor\\ (P21453)}}
& \multirow{7}{*}{\includegraphics[width=0.4\linewidth]{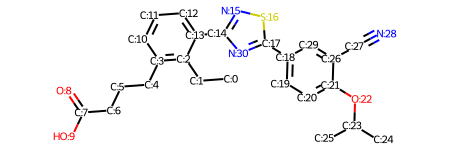}}
& & \multirow{7}{*}{\includegraphics[width=0.4\linewidth]{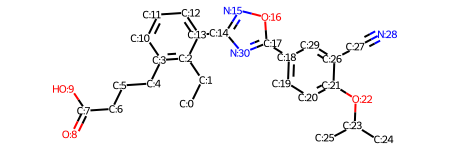}} & \\
&&&&Atom\#16: S$\rightarrow$O\\
&& Activity: 8.6002 && Activity: 9.6995 ($++$) \\
&& QED: 0.4766 && QED: 0.5177 ($+$) \\
&& SA: 2.7684 && SA: 2.5739 ($+$) \\
&& logP: 5.5007 && logP: 5.0322 ($+$) \\&&&&\\
\cmidrule(lr){2-5}
& \multirow{7}{*}{\includegraphics[width=0.4\linewidth]{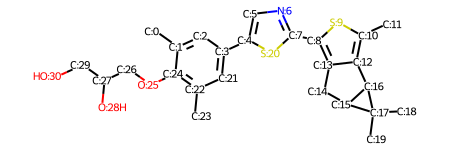}}
& & \multirow{7}{*}{\includegraphics[width=0.4\linewidth]{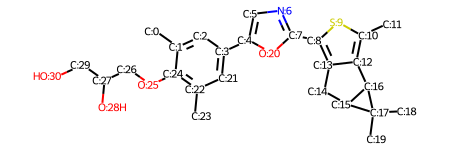}} & \\
&&&&Atom\#20: S$\rightarrow$O\\
&& Activity: 7.4225 && Activity: 9.0000 ($++$) \\
&& QED: 0.5158 && QED: 0.5638 ($+$) \\
&& SA: 4.1048 && SA: 4.1124 \\
&& logP: 5.4916 && logP: 5.0231 ($+$) \\&&&&\\
\cmidrule(lr){2-5}
& \multirow{7}{*}{\includegraphics[width=0.4\linewidth]{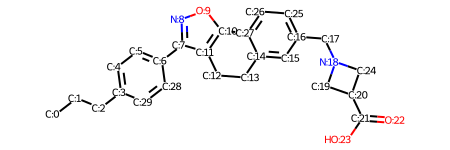}}
& & \multirow{7}{*}{\includegraphics[width=0.4\linewidth]{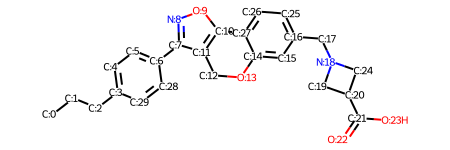}} & \\
&&&&Atom\#13: C$\rightarrow$O\\
&& Activity: 6.6576 && Activity: 8.8861 ($++$) \\
&& QED: 0.6539 && QED: 0.6575 \\
&& SA: 2.4926 && SA: 2.5408 \\
&& logP: 4.5761 && logP: 4.3699 ($+$) \\&&&&\\
\hline
\multirow{21}{*}{\tabincell{c}{Human\\orexin\\type 2\\receptor\\(O43614)}}
& \multirow{7}{*}{\includegraphics[width=0.4\linewidth]{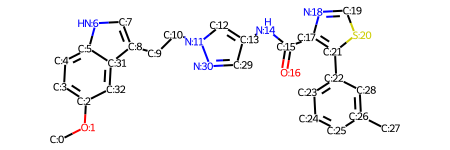}}
& & \multirow{7}{*}{\includegraphics[width=0.4\linewidth]{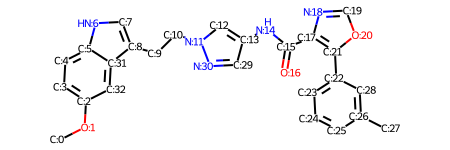}} & \\
&&&&Atom\#20: S$\rightarrow$O\\
&& Activity: 7.5229 && Activity: 8.6990 ($++$) \\
&& QED: 0.3473 && QED: 0.3769 ($+$) \\
&& SA: 2.6436 && SA: 2.6295 \\
&& logP: 5.2999 && logP: 4.8314 ($+$) \\&&&&\\
\cmidrule(lr){2-5}
& \multirow{7}{*}{\includegraphics[width=0.4\linewidth]{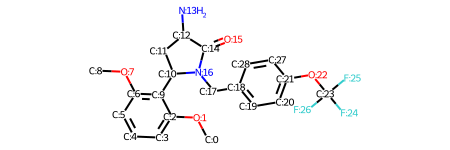}}
& & \multirow{7}{*}{\includegraphics[width=0.4\linewidth]{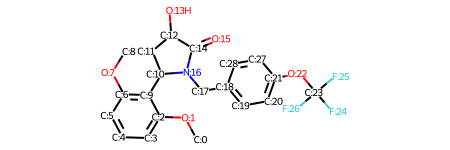}} & \\
&&&&Atom\#13: N$\rightarrow$O\\
&& Activity: 5.6882 && Activity: 7.2291 ($++$) \\
&& QED: 0.7903 && QED: 0.7889 \\
&& SA: 3.2084 && SA: 3.1571 \\
&& logP: 3.4033 && logP: 3.4369 \\&&&&\\
\cmidrule(lr){2-5}
& \multirow{7}{*}{\includegraphics[width=0.4\linewidth]{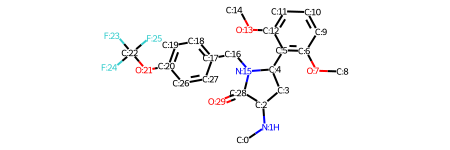}}
& & \multirow{7}{*}{\includegraphics[width=0.4\linewidth]{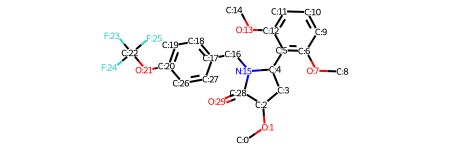}} & \\
&&&&Atom\#1: N$\rightarrow$O\\
&& Activity: 5.2358 && Activity: 6.7747 ($++$) \\
&& QED: 0.7352 && QED: 0.6690 ($-$) \\
&& SA: 3.2157 && SA: 3.2724 \\
&& logP: 3.6640 && logP: 4.0910 ($-$) \\&&&&\\
\hline
\multirow{21}{*}{\tabincell{c}{Human\\dopamine\\receptor\\ (P14416)}}
& \multirow{7}{*}{\includegraphics[width=0.4\linewidth]{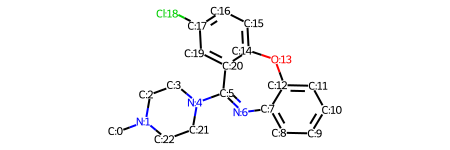}}
& & \multirow{7}{*}{\includegraphics[width=0.4\linewidth]{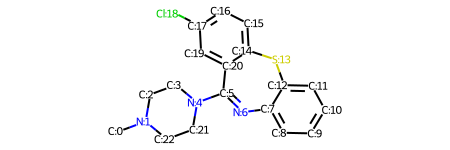}} & \\
&&&&Atom\#13: O$\rightarrow$S\\
&& Activity: 7.6778 && Activity: 9.3098 ($++$) \\
&& QED: 0.7370 && QED: 0.7164 ($-$) \\
&& SA: 2.2952 && SA: 2.4025 ($-$) \\
&& logP: 3.7714 && logP: 4.1303 ($-$) \\&&&&\\
\cmidrule(lr){2-5}
& \multirow{7}{*}{\includegraphics[width=0.4\linewidth]{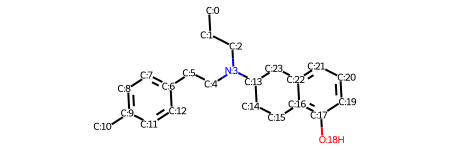}}
& & \multirow{7}{*}{\includegraphics[width=0.4\linewidth]{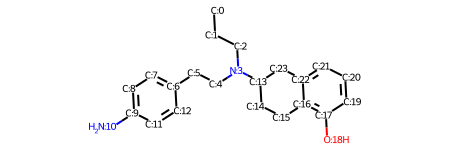}} & \\
&&&&Atom\#10: C$\rightarrow$N\\
&& Activity: 6.7696 && Activity: 8.1675 ($++$) \\
&& QED: 0.8451 && QED: 0.7950 ($-$) \\
&& SA: 2.6875 && SA: 2.7957 ($-$) \\
&& logP: 4.5126 && logP: 3.7864 ($+$) \\&&&&\\
\cmidrule(lr){2-5}
& \multirow{7}{*}{\includegraphics[width=0.4\linewidth]{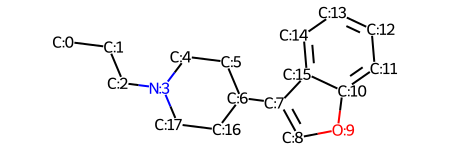}}
& & \multirow{7}{*}{\includegraphics[width=0.4\linewidth]{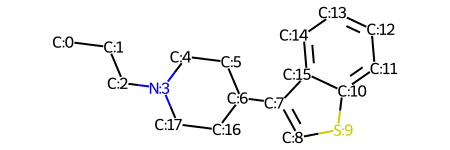}} & \\
&&&&Atom\#9: O$\rightarrow$S\\
&& Activity: 5.7545 && Activity: 7.1337 ($++$) \\
&& QED: 0.8091 && QED: 0.7847 ($-$) \\
&& SA: 2.1075 && SA: 2.0823 \\
&& logP: 4.0222 && logP: 4.4907 ($-$) \\&&&&\\
\hline
\end{tabular}}
\label{tGenHigh}
\end{table*}

\subsection{Generation of higher active drug molecules}
\label{secAO}
\emph{Setup:} To verify whether GAFSE can effectively generate matched high activity value drug molecules based on existing molecules, we screened out MMP-Cliffs (only one atom different, but their bioactivity values differ by more than 10 times), and the rest are used as the training set. We put molecules with lower activity values in MMP-Cliffs back into the training set to see if the molecules with higher activity values in MMP-Cliffs are included in our generated molecules.

\emph{Algorithm Review: }This section is about GAFSE experiments in the molecule optimization (MO) stage of lead discovery, referred to as GAFSE-MO. The pipeline of GAFSE-MO is as follows: First, molecules are input into the molecular embedding model $\operatorname{E}$ in the form of graphs to obtain molecular embeddings $\mathbf{f}$; then $\mathbf{f}$ is input to the downstream task model $\operatorname{N}$ to predict bioactivity values $\hat{y}$ and obtain the adversarial disturbance vector $\mathbf{d}$ according to Eq. \ref{eqAFSE}; next, $\mathbf{f}$ is input to the molecular reconstruction model $\operatorname{G}$ to reconstruct the initial graph features of input molecules $[\hat{\mathbf{a}}_{i},\hat{\mathbf{b}}_{i, j}]$; meanwhile, $\mathbf{f}+\mathbf{d}$ is input into $\operatorname{G}$ to get the optimized molecular graph features $[\widetilde{\mathbf{a}}_{i },\widetilde{\mathbf{b}}_{i,j}]$. In the training phase of GAFSE-MO, the output $\hat{y}$, $[\hat{\mathbf{a}}_{i},\hat{\mathbf{b}}_{i,j}] $ and $[\widetilde{\mathbf{a}}_{i},\widetilde{\mathbf{b}}_{i,j}]$ are optimized by Eq. \ref{eqGAFSE} to update the overall model parameters simultaneously. In the testing phase of GAFSE-MO, the output $\hat{y}$ is used to screen out highly active molecules (HS task), and the output $[\widetilde{\mathbf{a}}_{i},\widetilde {\mathbf{b}}_{i,j}]$ is used to optimize molecules.

\emph{Results:} Taking the orexin receptor O43614 in the \href{http://www.uniprot.org/docs/7tmrlist}{UniProt} database as an example, Figure \ref{t-sne} shows the location of GAFSE-generated molecules in the predicted structural feature-activity space. Results show that the predicted activities of most generated molecules are greatly increased, and their structural features are close to their original molecules, meeting the requirements of MMP-Cliffs. In Figure \ref{t-sne}, according to Eq. \ref{eqPsa}, the projection of solid arrows on the x-y plane indicates GAFSE's $\mathbf{d}$, which contains key information for generating MMP-Cliffs molecules. We queried the activity assays of orexin receptor O43614 and found that the true activities of four generated molecules (red stars in Figure \ref{t-sne}) are much higher than that of their original molecules. Besides, the molecule optimization results of GAFSE-MO can provide chemists with more insight into molecule optimization by analyzing the changes in elements at different atomic positions between the generated and original molecules (dotted arrows in Figure \ref{t-sne}).

\begin{table*}[t]
\centering
\caption{Comparison of classification indexes on ADMET benchmark datasets \cite{xiong2021admetlab}. Baseline results are taken from \cite{xiong2021admetlab}.}
\resizebox{\textwidth}{40mm}{
\begin{tabular}{cccccccccccc}
\hline
\multirow{2}[4]{*}{Category} & \multirow{2}[4]{*}{Model} & \multicolumn{2}{c}{AUC} & \multicolumn{2}{c}{ACC} & \multicolumn{2}{c}{MCC} & \multicolumn{2}{c}{Specificity} & \multicolumn{2}{c}{Sensitivity} \\
\cmidrule(lr){3-4}\cmidrule(lr){5-6}\cmidrule(lr){7-8}\cmidrule(lr){9-10}\cmidrule(lr){10-12} & & ADMETlab2.0 & GAFSE-MP & ADMETlab2.0 & GAFSE-MP & ADMETlab2.0 & GAFSE-MP & ADMETlab2.0 & GAFSE-MP & ADMETlab2.0 & GAFSE-MP \\
\hline
Absorption & Pgp-inhibitor & 0.922 & 0.876 & 0.867 & 0.737 & 0.723 & 0.504 & 0.844 & 0.583 & 0.882 & 0.897 \\
& Pgp-substrate & 0.840 & 0.848 & 0.768 & 0.703 & 0.538 & 0.456 & 0.705 & 0.947 & 0.828 & 0.444 \\
Distribution & BBB Penetration & 0.908 & 0.805 & 0.862 & 0.768 & 0.718 & 0.521 & 0.824 & 0.796 & 0.891 & 0.727 \\
Metabolism & CYP1A2 inhibitor & 0.928 & 0.662 & 0.852 & 0.670 & 0.704 & 0.364 & 0.848 & 0.479 & 0.857 & 0.857 \\
& CYP1A2 substrate & 0.737 & 0.864 & 0.649 & 0.800 & 0.298 & 0.611 & 0.632 & 0.734 & 0.667 & 0.875 \\
& CYP2C9 inhibitor & 0.919 & 0.732 & 0.841 & 0.725 & 0.671 & 0.361 & 0.823 & 0.538 & 0.878 & 0.815 \\
& CYP2C9 substrate & 0.725 & 0.914 & 0.707 & 0.860 & 0.386 & 0.710 & 0.776 & 0.822 & 0.606 & 0.886 \\
& CYP3A4 inhibitor & 0.921 & 0.897 & 0.832 & 0.832 & 0.659 & 0.654 & 0.825 & 0.752 & 0.841 & 0.891 \\
& CYP3A4 substrate & 0.776 & 0.983 & 0.713 & 0.961 & 0.437 & 0.918 & 0.820 & 0.957 & 0.608 & 0.966 \\
Toxicity & AMES Toxicity & 0.902 & 0.872 & 0.807 & 0.906 & 0.606 & 0.283 & 0.732 & 0.925 & 0.865 & 0.480 \\
& Eye Corrosion & 0.983 & 0.851 & 0.957 & 0.849 & 0.908 & 0.499 & 0.965 & 0.881 & 0.944 & 0.678 \\
& FDAMDD & 0.804 & 0.945 & 0.736 & 0.897 & 0.471 & 0.647 & 0.734 & 0.916 & 0.737 & 0.791 \\
& NR-AhR & 0.943 & 0.918 & 0.862 & 0.887 & 0.573 & 0.538 & 0.858 & 0.912 & 0.896 & 0.701 \\
& NR-AR-LBD & 0.915 & 0.885 & 0.936 & 0.958 & 0.472 & 0.560 & 0.942 & 0.964 & 0.783 & 0.783 \\
& NR-AR & 0.886 & 0.832 & 0.890 & 0.909 & 0.348 & 0.499 & 0.896 & 0.956 & 0.731 & 0.515 \\
& NR-Aromatase & 0.852 & 0.907 & 0.849 & 0.930 & 0.264 & 0.454 & 0.859 & 0.933 & 0.615 & 0.842 \\
& NR-ER-LBD & 0.850 & 0.901 & 0.903 & 0.913 & 0.364 & 0.465 & 0.918 & 0.924 & 0.618 & 0.714 \\
& NR-ER & 0.771 & 0.878 & 0.815 & 0.832 & 0.320 & 0.350 & 0.845 & 0.841 & 0.567 & 0.717 \\
& NR-PPAR-gamma & 0.893 & 0.918 & 0.896 & 0.839 & 0.344 & 0.678 & 0.901 & 0.828 & 0.750 & 0.852 \\
& Skin Sensitization & 0.707 & 0.905 & 0.775 & 0.833 & 0.462 & 0.653 & 0.539 & 0.819 & 0.889 & 0.863 \\
& SR-ARE & 0.863 & 0.905 & 0.827 & 0.817 & 0.469 & 0.624 & 0.850 & 0.830 & 0.701 & 0.798 \\
& SR-ATAD5 & 0.874 & 0.967 & 0.919 & 0.909 & 0.361 & 0.819 & 0.929 & 0.905 & 0.640 & 0.914 \\
& SR-HSE & 0.907 & 0.929 & 0.868 & 0.985 & 0.393 & 0.754 & 0.875 & 0.999 & 0.750 & 0.615 \\
& SR-MMP & 0.927 & 0.931 & 0.897 & 0.886 & 0.660 & 0.399 & 0.908 & 0.892 & 0.835 & 0.758 \\
& SR-p53 & 0.881 & 0.866 & 0.841 & 0.954 & 0.365 & 0.503 & 0.849 & 0.964 & 0.723 & 0.680 \\
\cmidrule(lr){2-12}
& Average & 0.865 & \textbf{0.880 } & 0.835 & \textbf{0.855 } & 0.501 & \textbf{0.553 } & 0.828 & \textbf{0.844 } & \textbf{0.764 } & \textbf{0.762 } \\
\hline
\end{tabular}}%
\label{tADMETc}%
\end{table*}%

\begin{table*}
\centering
\caption{Generation of matched non-toxic molecules from toxic molecules by GAFSE-MO.}
\resizebox{\textwidth}{75mm}{
\begin{tabular}{ccccc}
\hline
Targets & Anchors & Anchor Properties & Generated molecules & Generated Properties\\
\hline
\multirow{22.5}{*}{\tabincell{c}{NR-AhR Toxicity}}
& \multirow{7}{*}{\includegraphics[width=0.4\linewidth]{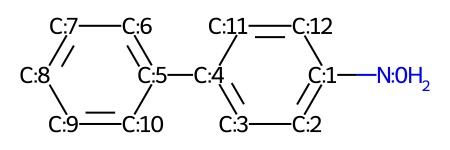}}
& & \multirow{7}{*}{\includegraphics[width=0.4\linewidth]{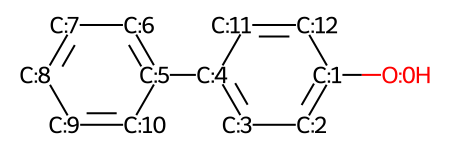}} & \\
&&&&Atom\#0: N$\rightarrow$O\\
&& Toxicity: High && Toxicity: Non-Toxic \\
&& QED: 0.6155 && QED: 0.6969 ($+$) \\
&& SA: 1.2681 && SA: 1.2254 \\
&& logP: 3.8570 && logP: 3.0592 ($+$) \\&&&&\\
\cmidrule(lr){2-5}
& \multirow{7}{*}{\includegraphics[width=0.4\linewidth]{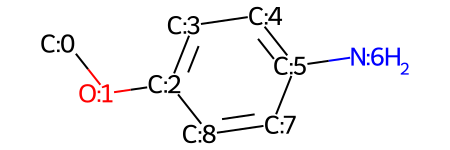}}
& & \multirow{7}{*}{\includegraphics[width=0.4\linewidth]{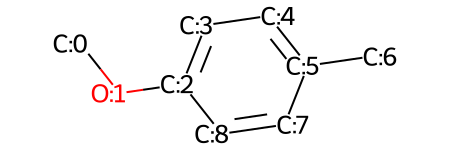}} & \\
&&&&Atom\#6: N$\rightarrow$C\\
&& Toxicity: High && Toxicity: Non-Toxic \\
&& QED: 0.5630 && QED: 0.5532 \\
&& SA: 1.4461 && SA: 1.0100 ($+$) \\
&& logP: 2.4220 && logP: 2.0036 \\&&&&\\
\cmidrule(lr){2-5}
& \multirow{7}{*}{\includegraphics[width=0.4\linewidth]{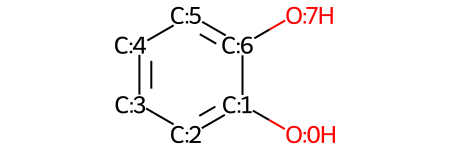}}
& & \multirow{7}{*}{\includegraphics[width=0.4\linewidth]{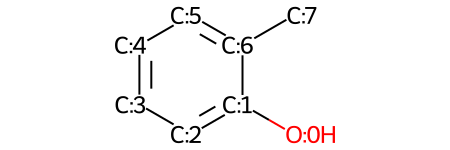}} & \\
&&&&Atom\#7: O$\rightarrow$C\\
&& Toxicity: High && Toxicity: Non-Toxic \\
&& QED: 0.3762 && QED: 0.5359 ($++$) \\
&& SA: 1.8671 && SA: 1.4050 ($+$) \\
&& logP: 1.4854 && logP: 1.7006 \\&&&&\\
\hline
\multirow{22.5}{*}{\tabincell{c}{NR-ER Toxicity}}
& \multirow{7}{*}{\includegraphics[width=0.4\linewidth]{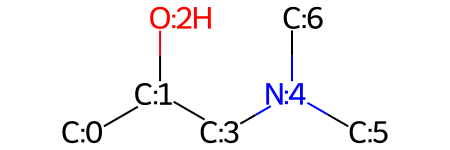}}
& & \multirow{7}{*}{\includegraphics[width=0.4\linewidth]{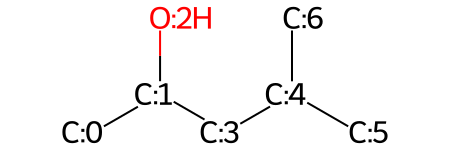}} & \\
&&&&Atom\#4: N$\rightarrow$C\\
&& Toxicity: High && Toxicity: Non-Toxic \\
&& QED: 0.5285 && QED: 0.5586 ($+$) \\
&& SA: 2.7983 && SA: 2.6752 \\
&& logP: -0.0712 && logP: 1.4133 ($++$) \\&&&&\\
\cmidrule(lr){2-5}
& \multirow{7}{*}{\includegraphics[width=0.4\linewidth]{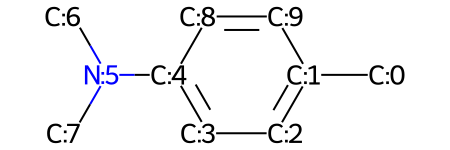}}
& & \multirow{7}{*}{\includegraphics[width=0.4\linewidth]{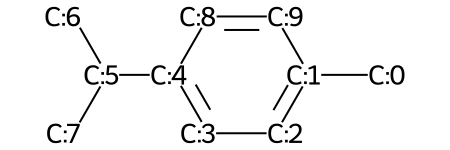}} & \\
&&&&Atom\#5: N$\rightarrow$C\\
&& Toxicity: High && Toxicity: Non-Toxic \\
&& QED: 0.5694 && QED: 0.5533 \\
&& SA: 1.3958 && SA: 1.2512 ($+$) \\
&& logP: 2.0610 && logP: 3.1184 ($-$)\\&&&&\\
\cmidrule(lr){2-5}
& \multirow{7}{*}{\includegraphics[width=0.4\linewidth]{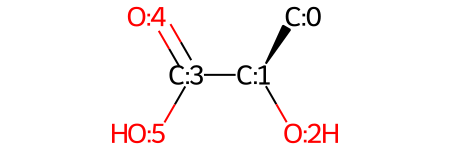}}
& & \multirow{7}{*}{\includegraphics[width=0.4\linewidth]{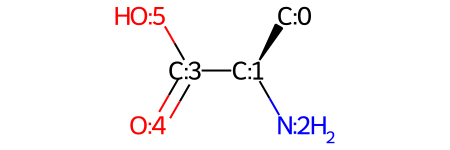}} & \\
&&&&Atom\#2: O$\rightarrow$N\\
&& Toxicity: High && Toxicity: Non-Toxic \\
&& QED: 0.4539 && QED: 0.4514 \\
&& SA: 2.3257 && SA: 2.3196 \\
&& logP: -0.5482 && logP: -0.5818 \\&&&&\\
\hline
\multirow{22.5}{*}{\tabincell{c}{AMES Toxicity}}
& \multirow{7}{*}{\includegraphics[width=0.4\linewidth]{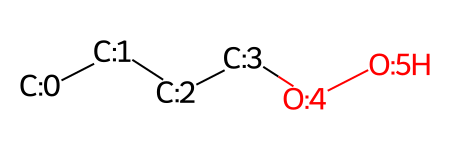}}
& & \multirow{7}{*}{\includegraphics[width=0.4\linewidth]{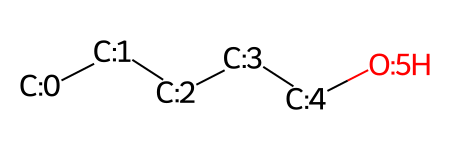}} & \\
&&&&Atom\#4: N$\rightarrow$C\\
&& Toxicity: High && Toxicity: Non-Toxic \\
&& QED: 0.3211 && QED: 0.5133 ($++$) \\
&& SA: 2.3758 && SA: 1.5665 ($+$) \\
&& logP: 1.2761 && logP: 1.1689 \\&&&&\\
\cmidrule(lr){2-5}
& \multirow{7}{*}{\includegraphics[width=0.4\linewidth]{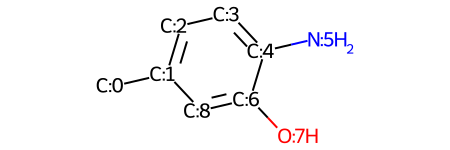}}
& & \multirow{7}{*}{\includegraphics[width=0.4\linewidth]{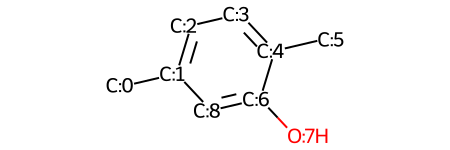}} & \\
&&&&Atom\#5: N$\rightarrow$C\\
&& Toxicity: High && Toxicity: Non-Toxic \\
&& QED: 0.4030 && QED: 0.5577 ($++$) \\
&& SA: 1.8514 && SA: 1.5860 ($+$) \\
&& logP: 1.2828 && logP: 2.0090 \\&&&&\\
\cmidrule(lr){2-5}
& \multirow{7}{*}{\includegraphics[width=0.4\linewidth]{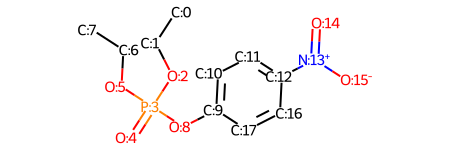}}
& & \multirow{7}{*}{\includegraphics[width=0.4\linewidth]{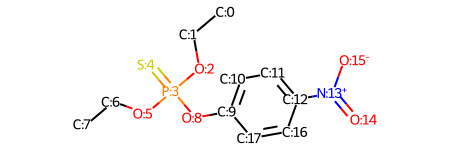}} & \\
&&&&Atom\#4: O$\rightarrow$S\\
&& Toxicity: High && Toxicity: Non-Toxic \\
&& QED: 0.4312 && QED: 0.4360 \\
&& SA: 2.3371 && SA: 2.5031 \\
&& logP: 3.1547 && logP: 3.2711 \\&&&&\\
\hline
\end{tabular}}
\label{tGenADMETTc}
\end{table*}

\begin{table*}
\centering
\caption{Generation of matched inhibitor molecules from non-inhibitor molecules by GAFSE-MO.}
\resizebox{\textwidth}{55mm}{
\begin{tabular}{ccccc}
\hline
Targets & Anchors & Anchor Properties & Generated molecules & Generated Properties\\
\hline
\multirow{14.5}{*}{\tabincell{c}{CYP1A2 Inhibitor}}
& \multirow{7}{*}{\includegraphics[width=0.4\linewidth]{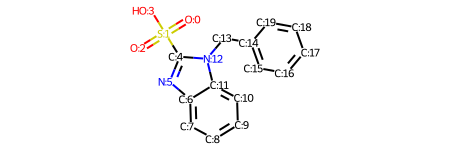}}
& & \multirow{7}{*}{\includegraphics[width=0.4\linewidth]{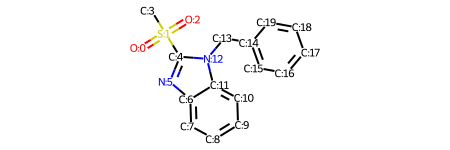}} & \\
&&&&Atom\#3: O$\rightarrow$C\\
&& Is Inhibitor: No && Is Inhibitor: Yes \\
&& QED: 0.7507 && QED: 0.7430 \\
&& SA: 1.9675 && SA: 1.8747 \\
&& logP: 2.3313 && logP: 2.4881 \\&&&&\\
\cmidrule(lr){2-5}
& \multirow{7}{*}{\includegraphics[width=0.4\linewidth]{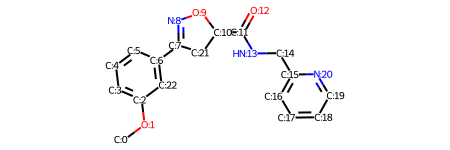}}
& & \multirow{7}{*}{\includegraphics[width=0.4\linewidth]{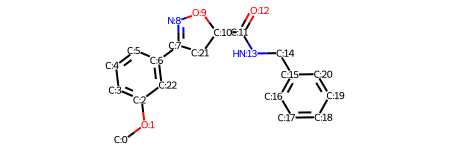}} & \\
&&&&Atom\#20: N$\rightarrow$C\\
&& Is Inhibitor: No && Is Inhibitor: Yes \\
&& QED: 0.9151 && QED: 0.9228 \\
&& SA: 2.6567 && SA: 2.4424 ($+$) \\
&& logP: 1.8996 && logP: 2.5046 \\&&&&\\
\hline
\multirow{14.5}{*}{\tabincell{c}{CYP2C9 Inhibitor}}
& \multirow{7}{*}{\includegraphics[width=0.4\linewidth]{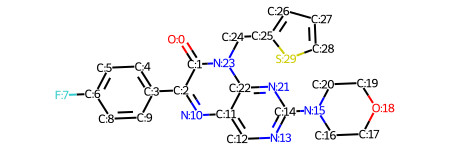}}
& & \multirow{7}{*}{\includegraphics[width=0.4\linewidth]{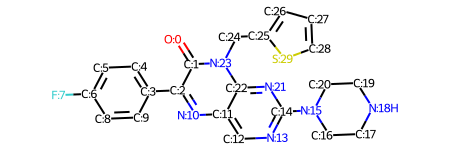}} & \\
&&&&Atom\#18: O$\rightarrow$N\\
&& Is Inhibitor: No && Is Inhibitor: Yes \\
&& QED: 0.5026 && QED: 0.5447 ($+$) \\
&& SA: 2.5278 && SA: 2.6015 \\
&& logP: 2.9390 && logP: 2.5120 ($+$) \\&&&&\\
\cmidrule(lr){2-5}
& \multirow{7}{*}{\includegraphics[width=0.4\linewidth]{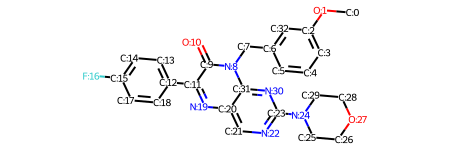}}
& & \multirow{7}{*}{\includegraphics[width=0.4\linewidth]{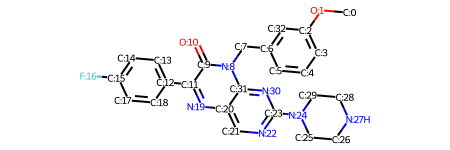}} & \\
&&&&Atom\#27: O$\rightarrow$N\\
&& Is Inhibitor: No && Is Inhibitor: Yes \\
&& QED: 0.4653 && QED: 0.5041 ($+$) \\
&& SA: 2.4246 && SA: 2.4922 \\
&& logP: 2.8861 && logP: 2.4591 \\&&&&\\
\hline
\multirow{14.5}{*}{\tabincell{c}{CYP3A4 Inhibitor}}
& \multirow{7}{*}{\includegraphics[width=0.4\linewidth]{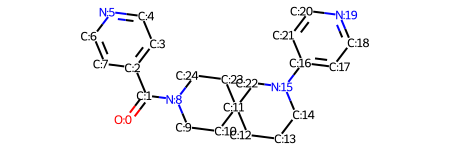}}
& & \multirow{7}{*}{\includegraphics[width=0.4\linewidth]{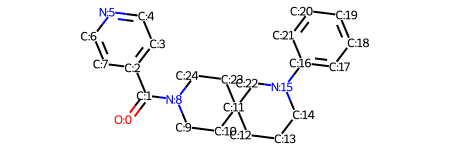}} & \\
&&&&Atom\#19: N$\rightarrow$C\\
&& Is Inhibitor: No && Is Inhibitor: Yes \\
&& QED: 0.8458 && QED: 0.8412 \\
&& SA: 2.8367 && SA: 2.6357 ($+$) \\
&& logP: 2.9994 && logP: 3.6044 ($-$) \\&&&&\\
\cmidrule(lr){2-5}
& \multirow{7}{*}{\includegraphics[width=0.4\linewidth]{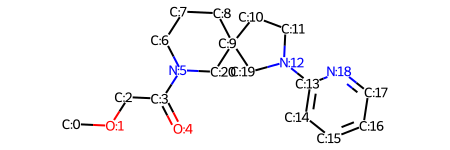}}
& & \multirow{7}{*}{\includegraphics[width=0.4\linewidth]{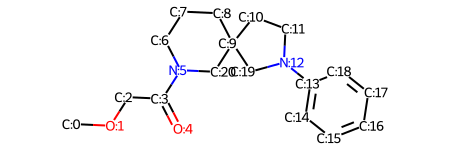}} & \\
&&&&Atom\#18: N$\rightarrow$C\\
&& Is Inhibitor: No && Is Inhibitor: Yes \\
&& QED: 0.8467 && QED: 0.8542 \\
&& SA: 3.2989 && SA: 3.1168 \\
&& logP: 1.5469 && logP: 2.1519 \\&&&&\\
\hline
\end{tabular}}
\label{tGenADMETMc}
\end{table*}

In Table \ref{tGenHigh}, we present representative cases of highly active molecules generation by GAFSE-MO. In the table, “Anchors” represent low activity molecules in the training set, and “Generated molecules” represent high activity molecules generated by GAFSE-MO based on “Anchors”. Results on multiple datasets show that the GAFSE-MO algorithm has a certain ability to generate highly active molecules in MMP-Cliffs. At the same time, we calculated various chemical properties of the generated molecules, including quantitative evaluation of drug-likeness (QED), synthesizable analysis (SA), and oil-water partition coefficient (LogP). Among them, the value of QED is between 0 and 1, and the larger the value, the higher the drug-likeness; the SA score is between 1 and 10, and the closer to 1, the easier synthesis; the drug is well absorbed when the logP is between 0 and 3, and smaller or larger values are not conducive to the absorption of the drug.

\subsection{ADMET Property Prediction}
\label{secADMETp}
\emph{Setup: } To verify the ability of GAFSE to accurately predict molecular ADMET properties, we evaluate the prediction performance of GAFSE on the available ADMET benchmark datasets constructed in \cite{xiong2021admetlab}. The experiments in this section use the same model settings as Section \ref{secScreening}, except $\lambda_1=0.08$ (Eq. \ref{eqGAFSE}) which is more suitable for multi-task learning. For a fair comparison with benchmark models of multi-task learning, we extend the multi-task learning scheme for GAFSE, that is, multi-tasks share model parameters other than the output layer. Among them, deeper neural networks with one more layer are used to learn multiple tasks, datasets with similar sample numbers are learned simultaneously, and the optimal classification threshold on the validation set is adopted. To reduce the computational cost, experiments in this section omit the molecule optimization part (Eq. \ref{eqGAFSE}), which is discussed in Section \ref{secPO}.

\emph{Algorithm Review: }This section is about GAFSE experiments in the molecular property prediction (MP) stage of lead discovery, referred to as GAFSE-MP. The difference between the algorithm of GAFSE-MP and GAFSE-HS in Section \ref{secScreening} is that the output prediction of the downstream task model $\operatorname{N}$ is molecular property values.

\emph{Results:} As shown in Table \ref{tADMETc}, compared to ADMETlab 2.0, GAFSE-MP achieves performance improvements on most metrics across all 25 ADMET classification tasks. On average, the area under the precision-recall curve (AUC) of GAFSE-MP increased by 1.73\%; the classification accuracy (ACC) increased by 2.40\%; the quality of the binary classification (Matthews correlation coefficient, MCC) improved by 10.38\%; and the specificity of binary classification is improved by 1.93\%. Although the sensitivity of binary classification is reduced by 0.26\% on average, these results can also show that the GAFSE framework can be used for classification and incorporate multi-task learning strategy, exhibiting its high scalability and potentiality to be studied as an open unified framework.

\subsection{ADMET Property Optimization}
\label{secPO}
\emph{Setup:} To verify whether GAFSE can effectively optimize ADMET properties of existing molecules, we screened out MMP-Cliffs (elements with only one atom are not the same, but the molecule changes from highly toxic to non-toxic, or from no specific properties to specific properties) in typical ADMET datasets in Table \ref{tADMETc}, and the rest are used as the training set. Then, we put the MMP-Cliffs molecules with strong toxicity or no specific properties back into the training set, and observe whether the MMP-Cliffs molecules generated by GAFSE have non-toxic or specific properties. The experiments in this section use the same model settings as Section \ref{secAO}.

\emph{Algorithm Review: }This section is about the GAFSE experiments in the molecule optimization (MO) stage of lead discovery, namely GAFSE-MO. The difference between GAFSE-MO in this section and Section \ref{secScreening} is that the downstream task model $\operatorname{N}$ predicts molecular property values, and the molecular reconstruction model $\operatorname{G}$ optimizes molecular properties.

\emph{Result:} In Table \ref{tGenADMETTc}, we present representative GAFSE-MO optimization results from highly toxic molecules to non-toxic MMP-Cliffs molecules. In the table, "Anchors" represent the highly toxic molecules in the training set, and "Generated molecules" represent the non-toxic molecules generated by GAFSE-MO based on "Anchors". Results on multiple datasets show that GAFSE-MO has a certain ability to optimize molecular toxicity. At the same time, we calculated various chemical properties of the resulting molecules, including quantitative evaluation of drug-likeness (QED), synthesizable analysis (SA), and oil-water partition coefficient (LogP). Among them, the value of QED is between 0 and 1, and the larger the value, the higher the drug-likeness; the SA is between 1 and 10, and the closer to 1, the easier synthesis; drugs with a logP between 0 and 3 are better absorbed, and smaller or larger values are less favorable for drug absorption. From Table \ref{tGenADMETTc}, it can be seen that most of the molecules have improved drug-like properties, higher synthesis feasibility, and better drug absorption after GAFSE toxicity optimization.

Additionally, Table \ref{tGenADMETMc} shows some examples of GAFSE-MO optimizing non-inhibitor molecules to inhibitor MMP-Cliffs molecules. In the table, "Anchors" refers to the non-inhibitor molecules in the training set, and "Generated molecules" refers to the inhibitor molecules generated by GAFSE-MO based on "Anchors". We also calculated the QED, SA, and LogP metrics for all molecules. It can be seen from the results that, compared with non-inhibitor molecules, GAFSE-optimized inhibitor molecules can become more drug-like, synthesizable, or absorbable.

\subsection{Exploring molecular generation on COVID-19}
\emph{Setup:} To explore the properties of GAFSE-optimized molecules in COVID-19-related databases, we used the AID1706 bioassay data in the PubChem database \footnote{https://pubchem.ncbi.nlm.nih.gov/bioassay/1706}, which is a high-throughput screening assay to identify inhibitors of the SARS coronavirus 3CLPro. For a fair comparison, we adopted the same experimental setup as \cite{li2022geometry}, that is, using 444 molecules with an activity score higher than 15 and less than 100 in a total of 290K molecules, 100 molecules were randomly selected as the test set, and the rest were used as the training set.

\begin{table}[t]
\centering
\caption{Comparison of molecule generation on AID1706. Baseline results are taken from \cite{li2022geometry}.}
\label{tAID}
\resizebox{\linewidth}{20mm}{
\begin{tabular}{c|c|c|c|c|c|c}
\hline
Model type & Model & Rec. & Val. & Uni. & Nov. & Avg. \\
\hline
\multirow{5}{*}{VAE-based} & JT-VAE & 76.7\% & 100\% & - & - & 88.35\%\\
& Character-VAE & 44.6\% & 0.7\% & - & - & 22.65\%\\
& Grammar-VAE & 53.7\% & 7.2\% & - & - & 30.45\%\\
& SD-VAE & 76.2\% & 43.5\% & - & - & 59.85\%\\
& GraphVAE & - & 13.5\% & - & - & 13.5\%\\
\hline
AR-based & AR-LSTM & - & 89.2\% & - & - & 89.2\%\\
\hline
\multirow{2}{*}{Flow-based} & GraphNVP & 100\% & 42.6\% & 94.8\% & \textbf{100\%} & 84.35\%\\
& GRF & \textbf{100\%} & 73.4\% & 53.7\% & \textbf{100\%} & 81.78\%\\
\hline
Geometry-based & GEOM-CVAE & \textbf{100\%} & 81.8\% & \textbf{100\%} & 94.11\% & 93.98\%\\
\hline
GAAE-based & GAFSE-MO & 94\% & \textbf{100\%} & \textbf{100\%} & \textbf{100\%} & \textbf{98.5\%}\\
\hline
\end{tabular}}
\end{table}

\begin{table}[t]
\centering
\caption{The comparison on the Top-5 QED scores of generated novel molecules. Baseline results are taken from \cite{li2022geometry}.}
\scalebox{0.87}{
\begin{tabular}{c|c|c|c|c|c}
\hline
Model & 1st & 2nd & 3rd & 4th & 5th \\
\hline
JT-VAE & 0.925 & 0.911 & 0.910 & - & - \\
\hline
GEOM-CVAE & 0.9442 & 0.9425 & 0.9120 & 0.9111 & 0.9089 \\
\hline
GAFSE-MO & \textbf{0.9461} & \textbf{0.9448} & \textbf{0.9363} & \textbf{0.9351} & \textbf{0.9241} \\
\hline
\end{tabular}}
\label{tAIDQED}
\end{table}

\begin{table}[t]
\centering
\caption{Generated novel molecules with the Top-5 QED scores by GAFSE-MO.}
\resizebox{\linewidth}{35mm}{
\begin{tabular}{cccc}
\hline
Rank & Canonical SMILES & Molecular graphs & Chemical Properties \\
\hline
\multirow{4}{*}{1st} & & \multirow{4}{*}{\includegraphics[width=0.3\linewidth]{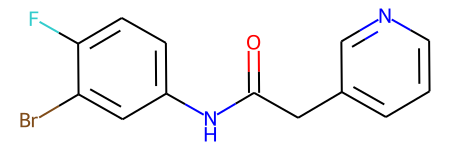}}& QED: 0.9461 ($++$)\\
& O=C(Cc1cccnc1) &&SA: 1.8498 ($++$) \\
& Nc1ccc(F)c(Br)c1 &&logP: 3.1644\\\\
\hline
\multirow{4}{*}{2rd} & & \multirow{4}{*}{\includegraphics[width=0.3\linewidth]{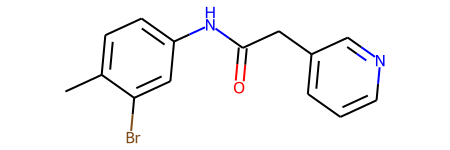}}& QED: 0.9448 ($++$)\\
& Cc1ccc(NC(=O) &&SA: 1.8236 ($++$) \\
& Cc2cccnc2)cc1Br &&logP: 3.3337\\\\
\hline
\multirow{4}{*}{3rd} & & \multirow{4}{*}{\includegraphics[width=0.3\linewidth]{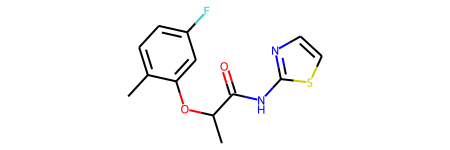}}& QED: 0.9363 ($++$)\\
& Cc1ccc(F)cc1OC &&SA: 2.5038 ($+$) \\
& (C)C(=O)Nc1nccs1 &&logP: 2.9966 ($+$)\\\\
\hline
\multirow{4}{*}{4th} & & \multirow{4}{*}{\includegraphics[width=0.3\linewidth]{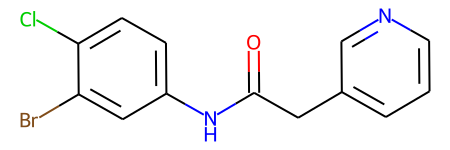}}& QED: 0.9351 ($++$)\\
& O=C(Cc1cccnc1) &&SA: 1.8990 ($++$) \\
& Nc1ccc(Cl)c(Br)c1 &&logP: 3.6787\\\\
\hline
\multirow{4}{*}{5th} & & \multirow{4}{*}{\includegraphics[width=0.3\linewidth]{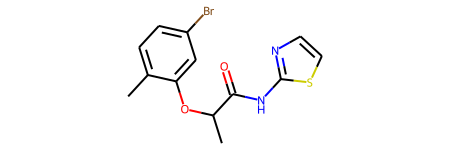}}& QED: 0.9241 ($++$)\\
& Cc1ccc(Br)cc1OC &&SA: 2.5552 ($+$) \\
& (C)C(=O)Nc1nccs1 &&logP: 3.6200\\\\
\hline
\end{tabular}}
\label{tAIDGen}
\end{table}

\emph{Algorithm Review: }This section is about GAFSE experiments in the molecule optimization (MO) stage of lead discovery, which is consistent with the GAFSE-MO algorithm in Section \ref{secAO}.

\emph{Evaluation metrics:} To comprehensively evaluate the performance of molecule generation, we use 4 commonly used metrics to test the model, namely reconstruction rate, validation rate, unique rate, and novelty rate. Among them, the reconstruction rate refers to the proportion of successfully reconstructed molecules in the test sets; the validation rate refers to the proportion of the molecules that meet the chemical specifications in the test sets; the unique rate refers to the proportion of unique molecules generated in the test set; the novelty rate refers to the proportion of generated molecules that differ from those in the test set.

\emph{Results:} In Table \ref{tAID}, we compare the performance of various types of molecular generative models, including (1) five variational autoencoder (VAE)-based models: junction tree (JT)-VAE \cite{jin2018junction}, Character-VAE \cite{gomez2018automatic}, Grammar-VAE \cite{kusner2017grammar}, syntax-directed (SD)-VAE \cite{dai2018syntax} and GraphVAE \cite{simonovsky2018graphvae}; (2) an atom-by-atom AR long short-term memory (LSTM) model: AR-LSTM \cite{li2018learning}; (3) two flow-base models: GraphNVP \cite{madhawa2019graphnvp} and graph residual flow (GRF) \cite{honda2019graph}; (4) a geometry-based constrained VAE model: GEOM-CVAE \cite{li2022geometry} and (5) graph adversarial autoencoder (GAAE)-based model: GAFSE-MO. Results in Table \ref{tAID} show that GAFSE-MO achieves 100\% in validation rate, unique rate, and novelty rate, and achieves a high reconstruction rate of 94\%, which significantly outperforms the current state-of-the-art molecular generation methods in average performance. The five molecules with the highest QED scores generated by GAFSE-MO and other methods are reported in Table \ref{tAIDQED}. It can be seen that the QED of GAFSE-generated molecules can reach higher. We present the SIMILES, molecular graphs, and chemical properties of these Top-5 molecules in Table \ref{tAIDGen}. Results show that the highly drug-like molecules generated by GFASE-MO based on the active molecules of COVID-19 are highly synthesizable and have good absorbability.

\newcolumntype{Z}[1]{>{\centering\let\newline\\\arraybackslash\hspace{0pt}}m{#1}}
\begin{table*}[!h]
\centering\small
\caption{Coherence Evaluation of Machine Learning Models across Lead Discovery Stages. Models are paired as compatible as possible.}
\begin{tabularx}{\textwidth}{Z{2cm}|Z{2cm}|Z{2cm}|Z{4.5cm}|Z{4.5cm}|Z{0.8cm}}
\hline
HS models & MP models & MO models & Compatibility & Incompatibility & Score \\
\hline
SVM, LR, $k$-NN, SEA \cite{unterthiner2014deep}, WDL-RF \cite{wu2018wdl} & RF \cite{svetnik2003random}, $k$-NN \cite{sakiyama2009use}, SVM \cite{heikamp2014support}, SwissADME \cite{daina2017swissadme}, CNN \cite{shi2019molecular}, ADMETlab \cite{dong2018admetlab}, admetSAR 2.0 \cite{yang2019admetsar} & All mentioned MO models & None & MP models’ Encoder: input molecular descriptors or images (CNN), no graph embedding output (-2); & $<$-2\\
\hline
DGL-based models & MTNN-GCN \cite{montanari2019modeling}, ADMETlab2.0 \cite{xiong2021admetlab} & DeltaDelta \cite{jimenez2019deltadelta} & Encoder: output graph embedding vectors (+1); Training Strategy: end-to-end (+1) & DeltaDelta’s Encoder: input paired protein and ligand voxelization, extra output protein embedding vectors (-3); DeltaDelta’s Evaluation Data: simulated or predicted values rather than experimentally measured (-1) & -2\\
\hline
DGL-based models & MTNN-GCN \cite{montanari2019modeling}, ADMETlab2.0 \cite{xiong2021admetlab} & MolEvol \cite{chen2021molecule} & Encoder: input one molecular graph, output graph embedding vectors (+1) & MolEvol’s Encoder: extra input a paired molecular graph (-1); MolEvol’s Training Strategy: two stages (-1); & -1\\
\hline
DGL-based models: GCNs \cite{kipf2016semi}, GATs \cite{velivckovic2017graph}, GINs \cite{xu2018powerful}, Neural FP \cite{duvenaud2015convolutional}, Weave \cite{kearnes2016molecular}, MPNNs \cite{gilmer2017neural}, Attentive FP \cite{xiong2019pushing}, RealVS \cite{yin2021realvs}, AFSE \cite{yin2022afse} & MTNN-GCN \cite{montanari2019modeling}, GNN models \cite{wieder2020compact}, ADMETlab2.0 \cite{xiong2021admetlab} & CORE \cite{fu2020core}, $\alpha$-MOP \cite{fu2020alpha}, Modof \cite{chen2021deep}, MOLER \cite{fu2021moler}, SPEAR \cite{fu2021spear}, SCVAE \cite{yu2022structure} & Encoder: input one molecular graph, output graph embedding vectors (+1); Training Strategy: end-to-end (+1) & CORE, $\alpha$-MOP, Modof, MOLER, SPEAR, and SCVAE’s Encoder: extra input a paired moelcular graph and their junction trees, extra output junction tree embedding vectors (-2); CORE, $\alpha$-MOP, Modof, SPEAR, and MOLER’s Evaluation Data: simulated or predicted values rather than experimentally measured (-1) & -1$\sim$0 \\
\hline
DGL-based models & MTNN-GCN \cite{montanari2019modeling}, GNN models \cite{wieder2020compact}, ADMETlab2.0 \cite{xiong2021admetlab} & UGMMT \cite{barshatski2021unpaired} & Encoder: input one molecular graph, output graph embedding vectors (+1); Training Strategy: end-to-end (+1) & UGMMT’s Encoder: extra input a paired molecular graph (-1); UGMMT’s Evaluation Data: simulated or predicted values rather than experimentally measured (-1) & 0\\
\hline
DGL-based models & MTNN-GCN \cite{montanari2019modeling}, ADMETlab2.0 \cite{xiong2021admetlab} & QMO \cite{hoffman2022optimizing} & Training Strategy: end-to-end (+1); Evaluation Data: assayed molecular properties through biological wet experiments (+1) & QMO’s Encoder: input molecular sequences (-1), output sequential embeddings & +1\\
\hline
\multicolumn{3}{c|}{GAFSE} & Encoder: input one molecular graph, output graph embedding vectors (+1); Training: end-to-end (+1); Evaluation Data: assayed molecular activities and properties through biological wet experiments (+2) & None & \textbf{\textbf{+4}} \\
\hline
\end{tabularx}
\label{tCoherency}
\end{table*}

\section{Discussion}
\subsection{Urgent Need for A One-Stop Framework across Lead Discover Stages}
To underscore the urgent need for a one-stop framework across lead discovery stages, we evaluate the compatibility of discrete models in their lead discovery stages (see Table \ref{tCoherency}), including hit screening (HS), molecular property prediction (MP) and molecule optimization (MO) models. When we try to unify these discrete models, we find that they suffer from various types of incompatibilities. For example, when the encoder is shared, the input and output are different, which leads to the failure of part of the encoder at different stages; different training strategies increase the difficulty of simultaneous convergence of the models across different stages; simulated or predicted molecular property values, rather than biological wet experimental values, are used to train and evaluate models, which hinders the widespread use of incorporated models. Therefore, as a one-stop framework, GAFSE effectively unifies all stages of lead discovery and is an important advance in improving the efficiency of drug developers and drug discovery success rates.

The following is an introduction to the models included in Table \ref{tCoherency}. HS models can be divided into two categories: deep graph learning (DGL)-based and non-DGL-based. DGL-based models include: graph convolutional networks (GCNs) \cite{kipf2016semi}, graph attention networks (GATs) \cite{velivckovic2017graph}, graph isomorphism networks (GINs) \cite{xu2018powerful} pre-trained with supervised learning and context prediction \cite{hu2019strategies}, neural fingerprint (Neural FP) \cite{duvenaud2015convolutional}, Weave \cite{kearnes2016molecular}, message passing neural networks (MPNNs) \cite{gilmer2017neural}, attentive fingerprint (Attentive FP) \cite{xiong2019pushing}, real virtual screening (RealVS) \cite{yin2021realvs}, and adversarial feature subspace enhancement (AFSE) \cite{yin2022afse}; Non-DGL-based models include: support vector machines (SVM), logistic regression (LR), $k$-nearest neighbor ($k$-NN), similarity ensemble approaches (SEA) \cite{unterthiner2014deep}, weighted deep learning combined with random forest (WDL-RF) \cite{wu2018wdl}; MP prediction models include: random forest (RF) \cite{svetnik2003random}, $k$-nearest neighbor ($k$-NN) \cite{sakiyama2009use}, support vector machines (SVM) \cite{heikamp2014support}, SwissADME \cite{daina2017swissadme}, ADMETlab \cite{dong2018admetlab}, admetSAR 2.0 \cite{yang2019admetsar}, fully-connected multitask networks and graph convolutional multitask networks (MTNN-GCN) \cite{montanari2019modeling}, convolutional neural network (CNN) \cite{shi2019molecular}, graph neural network models (GNN models) reviewed in \cite{wieder2020compact}, and ADMETlab 2.0 \cite{xiong2021admetlab}; MO models include: automatic molecule optimization using copy \& refine strategy (CORE) \cite{fu2020core}, molecule optimization with $\alpha$-divergence ($\alpha$-MOP) \cite{fu2020alpha}, DeltaDelta neural networks for lead optimization of small molecule potency (DeltaDelta) \cite{jimenez2019deltadelta}, molecule optimization by explainable evolution (MolEvol) \cite{chen2021molecule}, a deep generative model for molecule optimization via one fragment modification (Modof) \cite{chen2021deep}, molecule-level reward functions (MOLER) \cite{fu2021moler}, unpaired Generative Molecule-to-Molecule Translation for Lead Optimization (UGMMT) \cite{barshatski2021unpaired}, self-supervised post-training enhancer for molecule
optimization (SPEAR) \cite{fu2021spear}, structure-aware conditional variational auto-encoder (SCVAE) \cite{yu2022structure}, and a generic query-based molecule optimization (QMO) \cite{hoffman2022optimizing}.)

\subsection{Absence of novel MMP-Cliffs generation studies}
As the most basic and most valuable molecule optimization method\cite{bajorath2017representation}, MMP-Cliffs usually contains high structure-activity relationship (SAR) information \cite{dimova2014extraction,turk2017coupling}, which inspires pharmacists to discover and design high-efficiency molecules \cite{sushko2014prediction}. MMP-Cliffs are widely used in medicinal chemistry to study changes in compound properties, including biological activity, toxicity, environmental hazards, etc. \cite{leach2006matched}. Existing MMP-Cliffs analysis methods mainly answer the following three questions: how to identify MMP-Cliffs\cite{bajorath2017representation,hu2012mmp,stumpfe2020computational,hu2020introducing,lukac2017turbocharging,hu2012extending}, how to predict MMP-Cliffs\cite{sushko2014prediction,tamura2021interpretation,park2022acgcn,iqbal2021prediction, tamura2020ligand, horvath2016prediction}, and how to optimize molecules according to MMP-Cliffs \cite{turk2017coupling, dimova2013medicinal, stumpfe2014recent}. However, these methods are still limited to MMP-Cliffs in existing molecular libraries, and cannot generate novel MMP-Cliffs to open up the idea of molecule optimization. Therefore, in this paper, GAFSE generates novel MMP-Cliffs by modifying elements on one single atomic site (Figure \ref{framework}b), which achieves a substantial optimization of molecular properties and provides an important reference for drug discovery.

\subsection{Openness Check: Incorporating Multi-Task Learning into the GAFSE Framework}
In the molecular property prediction experiments (Section \ref{secADMETp}), we found that optimizing more tasks simultaneously improves the overall performance of GAFSE-MP. Due to the confidentiality \footnote{https://admetmesh.scbdd.com/resources/DA} of the data, we only obtained part of the dataset in ADMETlab 2.0 \cite{xiong2021admetlab}. Nevertheless, GAFSE-MP still achieves competitive overall performances incorporating multi-task learning (see Table \ref{tADMETc}). These results indicate that the GAFSE framework is open. Researchers can adapt GAFSE to custom scenarios by replacing or adding advanced methods to further improve the efficiency and success rate of drug discovery.

\section{Conclusion}
GAFSE unifies the discovery of lead compounds under an open learning framework, develops algorithms based on their consistency and uniqueness, and solves unique problems in each step. Extensive experimental results demonstrate that GAFSE can efficiently generate novel and highly active MMP-Cliffs molecules, and its performance in predicting activity values and ADMET properties is as good as the state-of-the-art methods. Nevertheless, there are remains many unresolved problems in lead compound discovery. Therefore, GAFSE can be extended as an open framework with open source code, convenient interfaces, and documentation for community participation and subsequent development.

In the future, we will use a large molecular database to pre-train models in the GAFSE framework, so that they can be adapted to various downstream tasks. In addition, the definitions of MMP-Cliffs are varied, and so, we will try to transform a substructure to build multi-objective optimization models of MMP-Cliffs.

\section*{Data Availability}
All datasets used in this paper can be downloaded from our GitHub repositories \footnote{https://github.com/Yueming-Yin/GAFSE/tree/main/Data}. Among them, ADMET’s datasets were downloaded from ADMETlab 2.0 \footnote{https://admetmesh.scbdd.com/resources/DA}.

\section*{Code Availability}
Code developed in this paper can be downloaded from our GitHub repositories \footnote{https://github.com/Yueming-Yin/GAFSE}.

\section*{Acknowledgments}
This work was supported in part by the National Natural Science Foundation of China (62071242, 61571233, 61901229, 61872198, and 61971216); the Natural Science Foundation of Jiangsu Province (BK20201378). Other contributors to this work: Li Hoi Yeung, Adams Wai Kin Kong, Yanxiang Zhu (Support).

\bibliographystyle{IEEEtran}
\bibliography{Manuscript_GAFSE}

\end{sloppypar}
\end{document}